\documentclass[11pt]{article}
\usepackage{epsfig}
\usepackage{amsfonts,amssymb}
\usepackage{amsmath}
\usepackage{bbm,bm}
\usepackage{cite}
\allowdisplaybreaks[1]
	
\usepackage{tikz}
\usetikzlibrary{shapes,arrows,shadows,automata,positioning}
\newdimen\nodeDist
\usepackage{comment}
\usepackage[normalem]{ulem} 

\usepackage[
  colorlinks=true,
  linkcolor={red!50!black},
  citecolor={blue!50!black},
  filecolor=black,
  urlcolor=black,
  breaklinks=true
]{hyperref}
\usepackage{orcidlink}

\def\dralgo{{\tt DRalgo}}
\newcommand{\DRalgoVersion}{{\tt v1.5.0}}

\usepackage{color}
\usepackage{algorithm,algpseudocode}

\algnewcommand{\algorithmiccall}{{\bf Call}}
\algdef{SE}[FOR]{Call}{EndCall}[1]
  {\algorithmiccall\ #1\ \{}
  {\}} 

\def\backtick{\char18}
\usepackage{listings}
\definecolor{codeblue}{rgb}{0.5,0.5,0.5}
\definecolor{codered}{rgb}{0.5,0.5,0.5}
\definecolor{codegreen}{rgb}{0.5,0.5,0.5}

\lstdefinestyle{output}{
  backgroundcolor=\color{gray!10},
  frame=single,
  rulecolor=\color{black},
  rulesepcolor=\color{gray!10},
  aboveskip=-1.2em,
  belowskip=1em,
  basicstyle=\ttfamily\footnotesize,
  keywordstyle=\ttfamily\footnotesize,
  commentstyle=\ttfamily\footnotesize,
  emphstyle=\ttfamily\footnotesize,
  stringstyle=\ttfamily\footnotesize,
}

\usepackage{caption}
\DeclareCaptionFormat{listingformat}{\colorbox{black}{\color{white}\thelstlisting}}
\DeclareCaptionFormat{outputformat}{\colorbox{black}{\color{white}{\small out}}}
\lstnewenvironment{outputlisting}[1][]{%
  \captionsetup[lstlisting]{format=outputformat}%
  \lstset{style=output,caption={out},#1}%
}{\addtocounter{lstlisting}{-1}}

\usepackage{slashed}

\newcommand{\gammaE}{{\gamma_\rmii{E}}}

\newcommand{\suii}[1]{b_{#1}} 
\newcommand{\suiii}[1]{c_{#1}} 

\newcommand{\scalind}[1]{s_{#1}}

\newcommand{\Nf}{N_{\rm f}}

\newcommand{\Nc}{N_{\rm c}}

\newcommand{\mE}{m_\rmii{E}}

\newcommand{\gE}{g_\rmii{E}}
\newcommand{\gM}{g_\rmii{M}}

\newcommand{\gs}{g_\rmi{s}}

\newcommand{\sumint}[1]{\hbox{$\sum$}\!\!\!\!\!\!\!\int_{#1}}

\def\lsi{\raise0.3ex\hbox{$<$\kern-0.75em\raise-1.1ex\hbox{$\sim$}}}
\def\gsi{\raise0.3ex\hbox{$>$\kern-0.75em\raise-1.1ex\hbox{$\sim$}}}

\newcommand{\nn}{\nonumber \\}

\newcommand{\rmi}[1]{{\mbox{\scriptsize #1}}}
\newcommand{\rmii}[1]{{\mbox{\tiny\rm{#1}}}}

\newcommand{\re}{\mathop{\mbox{Re}}}
\newcommand{\im}{\mathop{\mbox{Im}}}

\newcommand{\Tint}[1]{{\hbox{$\sum$}\!\!\!\!\!\!\!\int\,}_{\!\!\!\!\raise-0.9ex\hbox{$\scriptstyle{#1}$}}}
\newcommand{\Tinti}[1]{{{\Sigma}\!\!\!\!\raise0.3ex\hbox{$\int$}_\rmii{${#1}$}}}
\newcommand{\Tintip}[1]{{{\Sigma'}\!\!\!\!\!\raise0.3ex\hbox{$\int$}_\rmii{${#1}$}}}

\newcommand{\deltabar}{\raise-0.02em\hbox{$\bar{}$}\hspace*{-0.8mm}{\delta}}

\numberwithin{equation}{section}

\newcommand{\tr}{\mathrm{tr}}

\makeatletter \@addtoreset{equation}{section} \makeatother
\renewcommand{\theequation}{\arabic{section}.\arabic{equation}}
\makeatletter
\renewcommand\section{\@startsection{section}{1}{\z@}%
  {-5.5ex \@plus -1ex \@minus -.2ex}
  {2.3ex \@plus.2ex}%
  {\normalfont\large\bfseries}}
\renewcommand\subsection{\@startsection{subsection}{2}{\z@}%
  {-3.25ex\@plus -1ex \@minus -.2ex}%
  {1.5ex \@plus .2ex}%
  {\normalfont\normalsize\bfseries}}
\renewcommand\thesection{\@arabic\c@section}
\renewcommand\thesubsection{\thesection.\@arabic\c@subsection}
\renewcommand{\@seccntformat}[1]{%
  \csname the#1\endcsname.\hspace{1.0em}}
\makeatother

\begin{document}

\flushbottom

\begin{titlepage}
\begin{minipage}{.5\textwidth}
\flushright
\includegraphics[height=1.5cm]{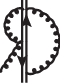}
\end{minipage}
\begin{minipage}{.5\textwidth}
\flushright
May 2026
\end{minipage}

\begin{centering}

\vfill

{\Large{\bf
Matching higher-dimensional operators\\
at finite temperature
for general models
}}

\vspace{0.8cm}

\renewcommand{\thefootnote}{\fnsymbol{footnote}}
Fabio Bernardo%
\orcidlink{0009-0008-0719-3219}%
,$^{\rm a,}$%
\footnote{fabio.bernardo@unige.ch}
Romain Guillermo Reinle%
\orcidlink{0009-0002-1057-4019}%
,$^{\rm b,a,}$%
\footnote{rreinle@student.ethz.ch}
and
Philipp Schicho%
\orcidlink{0000-0001-5869-7611} %
$^{\rm a,}$%
\footnote{philipp.schicho@unige.ch}

\vspace{0.8cm}

$^\rmi{a}$%
{\em
D\'epartement de Physique Th\'eorique, Universit\'e de Gen\`eve,
\\
24 quai Ernest Ansermet, CH-1211 Gen\`eve 4,
Switzerland\\}
\vspace{0.3cm}

$^\rmi{b}$%
{\em
Institute for Theoretical Physics, ETH Zurich, 8093 Z\"urich, Switzerland\\}
\vspace{0.3cm}

\vspace*{0.8cm}

\mbox{\bf Abstract}

\end{centering}

\vspace*{0.3cm}

\noindent
High-temperature dimensional reduction provides a systematic effective field theory framework
for studying finite-temperature thermodynamics and cosmological phase transitions.
While the matching of super-renormalizable operators
in the resulting three-dimensional effective theories is well established,
the matching of higher-dimensional operators has recently been reinvigorated.
These operators become phenomenologically relevant in strong first-order phase transitions
where they can quantify the convergence of the high-temperature expansion.
This work automates the matching of generic
three-dimensional dimension-five and -six operators
for arbitrary models containing scalars, fermions, and gauge fields,
implemented as an extension of the {\tt Mathematica} package \dralgo{}.
We present the operator basis, the matching procedure, and explicit examples
including a scalar-Yukawa model, hot QCD, and the full Standard Model up to dimension six,
covering
operators mixing the strong and electroweak sectors as well as parity-violating contributions.
Redundant operators, gauge dependence, and the corresponding field redefinitions
are discussed in detail.
The code and example model files are publicly available at
\url{https://github.com/DR-algo/DRalgo}.

\vfill
\end{titlepage}

{\hypersetup{hidelinks}
\tableofcontents
}
\clearpage

\renewcommand{\thefootnote}{\arabic{footnote}}
\setcounter{footnote}{0}

%
\clearpage

%
\section*{Program summary}
{\em Program title}: 
Dimensional Reduction algorithm (\dralgo{})
\\%
{\em Version}:
\DRalgoVersion
\\%
{\em Developer's repository link}:
\url{https://github.com/DR-algo/DRalgo}
\\%
{\em Licensing provisions}:
GNU General Public License 3 (GPLv3)
\\%
{\em Programming languages}:
{\tt Mathematica} 
\\%
{\em External routines/libraries}:
{\tt GroupMath}~\cite{ps.Fonseca:2020vke}
\\%
{\em Nature of problem}:
Construction of
high-temperature effective field theories for
beyond the Standard Model physics.
\\%
{\em Version updates}:
New functionalities for the matching of higher-dimensional operators\\%
{\em Solution method}:
Matching of $n$-point correlation functions using
tensor-notation of couplings~\cite{%
  ps.Martin:2017lqn,ps.Martin:2018emo,ps.Machacek:1984zw,
  ps.Machacek:1983fi,ps.Machacek:1983tz}.
\\%
{\em Restrictions}:
{\tt Mathematica} version 12 or above.

\clearpage

%
\section{Introduction}
\label{sec:intro}
Cosmological phase transitions provide a compelling window into
the physics of the early universe, with the potential to generate
observable signatures such as gravitational waves (GWs) from
strong first-order phase transitions.
These signals offer a unique opportunity to probe physics
beyond the Standard Model (BSM)~\cite{LISACosmologyWorkingGroup:2022jok}.

Effective field theory (EFT) techniques, 
in particular dimensional reduction at high temperatures~\cite{%
	Ginsparg:1980ef,Appelquist:1981vg,Nadkarni:1988fh,Landsman:1989be,
  Kajantie:1995dw,Braaten:1995jr,Braaten:1995cm},
play a central role in the systematic description of
the infrared (IR) dynamics~\cite{Linde:1980ts} of such transitions.
In this framework, the high-temperature dynamics of
a four-dimensional ($4d$) theory are mapped onto
a three-dimensional ($3d$) effective theory,
which can then be studied using both
perturbative and
non-perturbative methods~\cite{%
  Farakos:1994kx,Kajantie:1995kf,Kajantie:1996qd,Kajantie:1998yc};
see e.g.~\cite{%
  Niemi:2020hto,Kainulainen:2019kyp,Gould:2021dzl,Gould:2022ran}
for recent non-perturbative studies of phase transitions in BSM scenarios.

With their matching automated~\cite{Ekstedt:2022bff},
the leading contributions to
super-renormalizable operators in the $3d$~EFT are by now well understood.
The role of higher-dimensional operators, arising from subleading terms
in the high-temperature expansion, however,
remains an active area of research.
Recent studies have shown that such operators can play
a significant role in strong first-order phase transitions,
impacting both
non-perturbative phenomena~\cite{Chala:2025cya}
and the nature of the transition itself~\cite{%
  Kajantie:1995dw,Kajantie:1996mn,
  Bernardo:2025vkz,Chala:2024xll,Chala:2025xlk,Chala:2025aiz,
  Bernardo:2026whs}.
Strong transitions are of prime phenomenological interest
due to their enhanced stochastic gravitational wave background.

Including higher-dimensional operators significantly complicates
the matching procedure,
so that automation becomes necessary;
see e.g.~\cite{Fuentes-Martin:2022jrf,Carmona:2021xtq,Fuentes-Martin:2026bhr}.
We therefore perform
the dimensional reduction of a general class of theories relevant to
cosmological phase transitions, extending the matching procedure to include operators that,
in four space-time dimensions, would correspond to
dimension-five and dimension-six terms.
In the present work,
we restrict to contributions arising exclusively from
the hard scale $\pi T$.
Alternative approaches also integrate out lower scales,
such as the soft scale $g T$,
typically associated with the temporal scalar fields $A_0^a$
that arise from the zero components of the four-dimensional gauge fields.%
\footnote{%
  Thermal states break Lorentz boost symmetry,
  selecting the preferred rest frame of
  the plasma, $u_\mu = (1,0)$.
  The thermodynamics can nonetheless be formulated in a
  Lorentz-covariant way by including $u_\mu$
  in the tensor structures~\cite{Weldon:1982aq},
  which allows the temporal and spatial components
  $A_0^a$ and $A_i^a$ of the gauge field
  to renormalize independently
  without violating the Slavnov-Taylor identities.
}
Integrating out $A_0^a$ modes
thereby induces additional contributions to higher-dimensional operators.
However, it has been shown in~\cite{Bernardo:2025vkz} that such
approaches are not reliable in the case of strong phase transitions,
whereas dimensional reduction restricted to the hard scale can still provide
a valid description.

Utilizing these results,
we extend
the {\tt Mathematica} package \dralgo{}~\cite{Ekstedt:2022bff}
by automated higher-dimensional operator matching.
While \dralgo{} already handles the matching of
lower-dimensional, super-renormalizable operators in the three-dimensional EFT,
the resulting software allows for a systematic treatment of
higher-dimensional effects in general models.
Automated matching of higher-dimensional operators is also supported by
the package {\tt Matchotter}~\cite{Fuentes-Martin:2026bhr},
and the generic operator basis on
$\mathbb{R}^3 \times S^1$ has been derived in~\cite{Chakrabortty:2026swu}
with the matching procedure outlined in~\cite{Chakrabortty:2024wto}.
The present work details the \dralgo{} extension, providing explicit examples and
the complete results of the Standard Model (SM) dimensional reduction up
to dimension-six operators.

This paper is organized as follows.
In sec.~\ref{sec:setup}, we present
the dimension-five and dimension-six operators arising in
the three-dimensional effective theory after
dimensional reduction and integrating out the hard scale $\pi T$.
These operators are not present in
the original four-dimensional Lagrangian, but are generated radiatively
in the effective description.
We show how their Wilson coefficients are determined by
the parameters of the underlying four-dimensional theory,
and introduce the new \dralgo{} functionality that automates this matching.
Sec.~\ref{sec:application_to_models}
illustrates the use of these features
through three explicit model examples:
the scalar-Yukawa model,
QCD, and
the Standard Model.
Sec.~\ref{sec:details}
discusses technical aspects of the computation and gives guidance on
the use of the new features.
Sec.~\ref{sec:conclusions} summarizes our findings and outlines future directions.
Appendix~\ref{appendix:SM} contains
the complete set of dimensionally reduced Standard Model operators up to dimension six.
Appendix~\ref{appendix:QCD}
collects the operator basis and field redefinitions for hot QCD.

%
\section{Setup}
\label{sec:setup}

The most general Euclidean four-dimensional Lagrangian
is constructed from
scalar fields,
gauge fields,
two components Weyl fermions~\cite{Dreiner:2008tw}, and
ghost fields.
After defining the Euclidean Pauli matrices as
$\sigma^i \equiv i \sigma_\rmii{M}^i$,
where $\sigma_\rmii{M}^i$ are the Minkowskian Pauli matrices,
the Euclidean Weyl matrices read
\begin{align}
  \sigma_{\mu} &=\left(\mathbbm{1},\,\sigma^i\right)
  \,,&
  \overline{\sigma}_{\mu} &=\left(\mathbbm{1},\,-\sigma^i\right)
  \;,
\end{align}
which satisfy
$\sigma_\mu\overline{\sigma}_\nu + \sigma_\nu\overline{\sigma}_\mu = 2\delta_{\mu\nu}$.
The Lagrangian can be expressed in
the compact tensor notation~\cite{%
  Braaten:1995cm,Martin:2017lqn,Martin:2018emo,Ekstedt:2022bff}
\begin{align}
\label{eq:Lag_4d_DRalgo}
\mathcal{L}^{(4d)}&=
   \frac{1}{2} R_{\scalind{1}}^{ }(-\delta_{\scalind{1}\scalind{2}}^{ }\partial_\mu^{ }\partial_\mu^{ } + \mu_{\scalind{1}\scalind{2}}^{ }) R_{\scalind{2}}^{ }
  + \frac{1}{4} \delta_{ab}^{ } F_{\mu \nu}^{a} F_{\mu \nu}^{b} 
  + \frac{1}{2\xi_a}(\partial_\mu A_{\mu}^{a})^2
  + \partial_\mu \overline{\eta}^a \partial_\mu \eta^a
  \nn &
  + \psi_{I}^{\dagger}\overline{\sigma}_\mu^{ } \partial_\mu^{ } \psi_{I}^{ }
  + \frac{1}{2}(M_{IJ}\psi_{I}^{ }\psi_{J}^{ }+\mathrm{h.c.})
  + \mathcal{L}^{(4d)}_{\rmi{int}}
  \;,\\[3mm]
\mathcal{L}^{(4d)}_{\rmi{int}}&=
    \lambda_{\scalind{1}}^{ }R_{\scalind{1}}^{ }
  + \frac{1}{3!}\lambda_{\scalind{1}\scalind{2}\scalind{3}}^{ } R_{\scalind{1}}^{ } R_{\scalind{2}}^{ } R_{\scalind{3}}^{ }
  + \frac{1}{4!}\lambda_{\scalind{1}\scalind{2}\scalind{3}\scalind{4}}^{ } R_{\scalind{1}}^{ } R_{\scalind{2}}^{ } R_{\scalind{3}}^{ } R_{\scalind{4}}^{ }
  + \frac{1}{2}(Y_{\scalind{1} I J} R_{\scalind{1}}^{ } \psi_{I}^{ } \psi_{J}^{ } + \mathrm{h.c.})
  \nn &
  + ig_{IJ}^{a}A^a_\mu \psi_{I}^{\dagger}\overline{\sigma}_\mu^{ } \psi_{J}^{ }
  + g_{\scalind{1}\scalind{2}}^{a} A^a_\mu  R_{\scalind{1}}^{ } \partial_\mu^{ } R_{\scalind{2}}^{ }
  + \frac{1}{2}g^{a}_{\scalind{1} \scalind{3}}g^b_{\scalind{2} \scalind{3}}A_\mu^a A_{\mu}^{b} R_{\scalind{1}}^{ } R_{\scalind{2}}^{ }
  + g^{abc}A_{\mu}^{a}A_{\nu}^{b}\partial_\mu^{ } A^c_\nu
  \nn &
  + \frac{1}{4}g^{abe}g^{cde}A_{\mu}^{a}A_{\nu}^{b}A_{\mu}^{c} A_{\nu}^{d}
  - g^{abc}A_\mu^a \eta^b \partial_\mu^{ } \overline{\eta}^c
  \,,
\end{align}
where $\mathcal{L}^{(4d)}_\rmi{int}$ collects all interaction terms, 
$F_{\mu\nu} = \frac{i}{g} [D_\mu, D_\nu]$
is the field-strength tensor, and
$\xi_a$ are the gauge-fixing parameters
related to $A^a_\mu$
in the covariant gauge~\cite{Martin:2018emo}.
The real scalar fields $R_s$ carry scalar indices $s\in \{\scalind{1},\dots,\scalind{n}\}$;
in the Standard Model these correspond to the Higgs, the neutral Goldstone,
and the real and imaginary components of the charged Goldstone.
The real gauge fields $A^a_\mu$ carry gauge indices $\{a,b,c,d\}$,
$\eta^a$ denotes the ghost fields, and
$\psi_I$ is a Weyl spinor with fermion indices $\{I,J,K,L\}$.
The parameters $\mu_{\scalind{1}\scalind{2}}$ are the (squared) scalar mass matrix entries
and $M_{IJ}$ the fermion mass matrix entries.
Repeated indices are summed over regardless of their vertical placement.
The coupling tensors appearing in $\mathcal{L}^{(4d)}_\rmi{int}$
correspond to the implementation in \dralgo{}~\cite{Ekstedt:2022bff}
as listed in tab.~\ref{tab:coupling_tensors}.
\begin{table}[t]
  \centering
\begin{tabular}{| r | c | l |} 
  \hline
  \dralgo{} &
  Lagrangian &
  Description \\
  \hline
  \hline
  {\tt gvvv} & $g^{abc}$
  & group structure constants \\ 
  {\tt gvff} & $g_{IJ}^{a}$
  & vector-fermion trilinear couplings \\
  {\tt gvss} & $g_{\scalind{1}\scalind{2}}^{a}$
  & vector-scalar trilinear couplings \\ 
  {\tt $\lambda$1} & $\lambda_{\scalind{1}}$
  & scalar tadpole couplings\ \\
  {\tt $\lambda$3} & $\lambda_{\scalind{1}\scalind{2}\scalind{3}}$
  & scalar cubic couplings \\
  {\tt $\lambda$4} & $\lambda_{\scalind{1}\scalind{2}\scalind{3}\scalind{4}}$
  & scalar quartic couplings \\
  {\tt $\mu$ij} & $\mu_{\scalind{1}\scalind{2}}$
  & scalar-mass matrix \\
  {\tt $\mu$IJ}, 
  {\tt $\mu$IJC} &
  $M_{IJ}$,
  $M^{IJ}$
  & fermion-mass matrices \\ 
  {\tt Ysff}, 
  {\tt YsffC} &
  $Y_{\scalind{1}IJ}$,
  $Y_{\scalind{1}}^{IJ}$
  & Yukawa couplings \\
  \hline
\end{tabular}
  \caption{%
    Coupling tensors in the four-dimensional Euclidean Lagrangian~\eqref{eq:Lag_4d_DRalgo}
    and their corresponding input symbols in \dralgo{}~\cite{Ekstedt:2022bff}.
  }
  \label{tab:coupling_tensors}
\end{table}

The generic Lagrangian in eq.~\eqref{eq:Lag_4d_DRalgo} includes
all operators up to mass dimension four~\cite{Ekstedt:2022bff}.
In general the four-dimensional theory may also contain higher-dimensional operators,
such as a sextic term $R_{\scalind{1}} \dots R_{\scalind{6}}$
generated from ultraviolet (UV) physics at
a high scale $\Lambda_\rmii{UV} \gg T$~\cite{Aebischer:2025qhh}.
While such operators in the context of SMEFT,
have been dimensionally reduced in~\cite{Kajantie:1995dw,Croon:2020cgk,Chala:2025xlk},
our analysis will not include such operators.

Henceforth, we focus on higher-dimensional operators
obtained via dimensional reduction from the hard to the soft scale,
by integrating out modes of $\mathcal{O}(\pi T)$.
Future extensions of the software may incorporate
the soft scale $(gT)$,
in particular the temporal scalars $A_0$.
See~\cite{Giovannangeli:2005rz,KorthalsAltes:2017jzj,Laine:2018lgj}
for examples in QCD.
In phenomenologically relevant phase-transition scenarios,
however,
integrating out the soft scale $gT$
cannot resolve the transition dynamics~\cite{%
  Kajantie:1998yc,Vuorinen:2006nz,Chakrabortty:2024wto,Bernardo:2025vkz}.
Strong transitions can be driven by scales $\Lambda > gT$,
rendering integrating out the soft-scale unreliable.
By contrast, integrating out only the hard scale $\pi T$
remains justified within certain regions of parameter space.

We derive the matching relations for the following
dimensionally reduced $3d$ effective operator basis%
\footnote{%
		These operators correspond
    to dimensions five and six in four space-time dimensions ($4d$).
}
adapted from~\cite{Fonseca:2025zjb}.
Additionally,
the temporal scalar fields $A_0$ are made explicit and 
$3d$ parity-violating operators~\cite{Kajantie:1997ky} are incorporated.
Here, we display
the dimension-five operators,
\begin{align}
\label{eq:dim_5_operators}
	\mathcal{L}^{(3d)}_5&=
			\frac{1}{2}O^{(5,1)}_{\scalind{1}a_2a_3}F^{a_2}_{ij} F^{a_3}_{ij} R_{\scalind{1}}^{ }
		+ \frac{1}{5!}O^{(5,2)}_{\scalind{1}\scalind{2}\scalind{3}\scalind{4}\scalind{5}} R_{\scalind{1}}^{ } R_{\scalind{2}}^{ } R_{\scalind{3}}^{ } R_{\scalind{4}}^{ } R_{\scalind{5}}^{ }
		\nn &
		+ \frac{1}{3!2!}O^{(5,3)}_{\scalind{1}\scalind{2}\scalind{3}a_4a_5}
    R_{\scalind{1}}^{ } R_{\scalind{2}}^{ } R_{\scalind{3}}^{ } A_0^{a_4}A_0^{a_5}
		+ \frac{1}{4!}O^{(5,4)}_{\scalind{1}a_2a_3a_4a_5}R_{\scalind{1}}^{ }A_0^{a_2}A_0^{a_3}A_0^{a_4}A_0^{a_5}
		\nn &
		+ \frac{1}{2!}O^{(5,5)}_{\scalind{1}\scalind{2}\scalind{3}}(D_jD_jR)_{\scalind{1}}R_{\scalind{2}}^{ }R_{\scalind{3}}^{ }
		+ \frac{1}{2!}O^{(5,6)}_{\scalind{1}a_2a_3}(D_jD_jR)_{\scalind{1}}A_0^{a_2}A_0^{a_3}
		\nn &
		+ O^{(5,7)}_{\scalind{1}a_2a_3}(D_jD_jA_0)^{a_2}A_0^{a_3}R_{\scalind{1}}^{ }
		+ O^{(5,8)}_{\scalind{1}a_2a_3}R_{\scalind{1}}^{ }(D_j^{ } A_0)^{a_2}\widetilde{F}_j^{a_3}
    \,,
\end{align}
and the corresponding dimension-six operators,
\begin{align}
\label{eq:dim_6_operators}
    \mathcal{L}^{(3d)}_6 &=
			\frac{1}{3!}O^{(6,1)}_{a_1 a_2 a_3}F^{a_1}_{ij} F^{a_2}_{jk} F^{a_3}_{ki}
		+ \frac{1}{2!2!}O^{(6,2)}_{\scalind{1} \scalind{2} a_3 a_4}\,F^{a_3}_{ij}F^{a_4}_{ij}R_{\scalind{1}}^{ }R_{\scalind{2}}^{ }
		\nn &
		+ \frac{1}{2!2!}O^{(6,3)}_{a_1 a_2 a_3 a_4}\,F^{a_1}_{ij}F^{a_2}_{ij}A_{0}^{a_3}A^{a_4}_0
		+ \frac{1}{2!2!}O^{(6,4)}_{\scalind{1} \scalind{2} \scalind{3} \scalind{4}}(D_i^{ }R)_{\scalind{1}}(D_i^{ }R)_{\scalind{2}}R_{\scalind{3}}R_{\scalind{4}}
		\nn &
		+ \frac{1}{2!2!}O^{(6,5)}_{\scalind{1} \scalind{2} \scalind{3} \scalind{4}}(D_i^{ }R)_{\scalind{1}}(D_i^{ }R)_{\scalind{2}}A^{a_3}_0A^{a_4}_0
		+ O^{(6,6)}_{\scalind{1} \scalind{2} a_3 a_4}(D_i^{ }A_0)^{a_3}(D_i^{ }R)_{\scalind{1}}A^{a_4}_0R_{\scalind{2}}
		\nn &
		+ \frac{1}{2!2!}O^{(6,7)}_{\scalind{1} \scalind{2} a_3 a_4}(D_i^{ }A_0)^{a_3}(D_i^{ }A_0)^{a_4}R_{\scalind{1}}R_{\scalind{2}}
        + \frac{1}{2!2!}O^{(6,8)}_{a_1 a_2 a_3 a_4}(D_i^{ }A_0)^{a_1}(D_i^{ }A_0)^{a_2}A^{a_3}_0A^{a_4}_0
		\nn &
		+ \frac{1}{6!}\,O^{(6,9)}_{\scalind{1}\scalind{2}\scalind{3}\scalind{4}\scalind{5}\scalind{6}}R_{\scalind{1}}^{ }R_{\scalind{2}}^{ }R_{\scalind{3}}^{ }R_{\scalind{4}}^{ }R_{\scalind{5}}^{ }R_{\scalind{6}}^{ }
		+ \frac{1}{4!2!}\,O^{(6,10)}_{\scalind{1} \scalind{2} \scalind{3} \scalind{4} a_5 a_6}R_{\scalind{1}}^{ }R_{\scalind{2}}^{ }R_{\scalind{3}}^{ }R_{\scalind{4}}^{ }B^{a_5}_0B^{a_6}_0
		\nn &
		+ \frac{1}{4!2!}\,O^{(6,11)}_{\scalind{1} \scalind{2} a_3 a_4 a_5 a_6}
      R_{\scalind{1}}^{ }R_{\scalind{2}}^{ }A^{a_3}_0A^{a_4}_0A^{a_5}_0A^{a_6}_0
		+ \frac{1}{6!}\,O^{(6,12)}_{a_1 a_2 a_3 a_4 a_5 a_6}
      A^{a_1}_0A^{a_2}_0A^{a_3}_0A^{a_4}_0A^{a_5}_0A^{a_6}_0
		\nn &
		+ \frac{1}{2!}O^{(6,13)}_{a_1 a_2}(D_i^{ } F_{i j})^{a_1}(D_k^{ } F_{k j})^{a_2}
		+ \frac{1}{2}O^{(6,14)}_{\scalind{1} \scalind{2} a_3}(D_j^{ } F_{i j})^{a_3}[(D_i^{ }R)_{\scalind{1}}R_{\scalind{2}}-(D_i^{ }R)_{\scalind{2}}R_{\scalind{1}}]
		\nn &
		+ \frac{1}{2!}O^{(6,15)}_{\scalind{1} \scalind{2}}(D_i^{ }D_i^{ }R)_{\scalind{1}}(D_j^{ }D_j^{ }R)_{\scalind{2}}
		+ \frac{1}{2}O^{(6,16)}_{a_1 a_2 a_3}(D_j^{ } F^{a_1}_{i j})[(D_i^{ }A_{0})^{a_2}A^{a_3}_0-(D_i^{ }A_{0})^{a_3}A^{a_2}_0]
		\nn &
		+ \frac{1}{2!}O^{(6,17)}_{a_1 a_2}(D_i^{ }D_i^{ }A_0)^{a_1}(D_j^{ }D_j^{ }A_0)^{a_2}
    + O^{(6,18)}_{a_1 a_2 a_3}A_0^{a_1}\widetilde{F}_k^{a_2}(D_l^{ }F_{lk}^{ })^{a_3}
		\nn &
		+ O^{(6,19)}_{\scalind{1} \scalind{2} a_3 a_4}R_{\scalind{1}}^{ }(D_i^{ }R)_{\scalind{2}}^{ }A_0^{a_3}\widetilde{F}_{i}^{a_4}
		+ \frac{1}{2!}O^{(6,20)}_{a_1 a_2 a_3 a_4}A_0^{a_1}A_0^{a_2}(D_i^{ }A_0)^{a_3}\widetilde{F}_i^{a_4}
    \,,
\end{align}
where
indices $i,j$ are related to the three-dimensional spatial coordinates, and
$\widetilde{F}^{a}_i \equiv \frac{1}{2}\epsilon_{ijk}^{ }F^{a}_{jk}$ is
the dual field-strength tensor.
The covariant derivatives for generic gauge and scalar fields
are defined as
\begin{align}
    (D_i^{ } R)_{\scalind{1}}&=
            \partial_{i}^{ }R_{\scalind{1}}^{ }
        -g^{a_1}_{\scalind{1}\scalind{2}}A^{a_1}_i R_{\scalind{2}}^{ }
    \,,\\[1mm]
    (D_i^{ } A_j^{ })^{a_1}&=
        \partial_i^{ } A_j^{a_1}
        +g^{a_1a_2a_3}A^{a_2}_i A^{a_3}_j
    \,.
\end{align}

The operator basis introduced above remains redundant,
since some operators are proportional to the equations of
motion of the fields.
For instance, the scalar and gauge equations of motion (EOMs),
\begin{align}
  \label{eq:scalar:EOM}
  D_i^{ } D_i^{ } R_{\scalind{1}} &=
      \mu_{\scalind{1} \scalind{2}}^{ } R_{\scalind{2}}^{ }
    + \frac{1}{2}\lambda_{\scalind{1} \scalind{2} \scalind{3}} R_{\scalind{2}}^{ } R_{\scalind{3}}^{ }
    + \dots
  \,,\nn[1mm]
  (D_i^{ } F_{ij})^{a_1} &=
      g^{a_1}_{\scalind{1} \scalind{2}}(D_j^{ } R_{\scalind{1}}) R_{\scalind{2}}^{ }
    + \dots
  \,,
\end{align}
allow operators such as
$O^{(6,15)}_{\scalind{1}\scalind{2}}(D_i^{ }D_i^{ }R)_{\scalind{1}}(D_j^{ }D_j^{ }R)_{\scalind{2}}$
or
$O^{(6,13)}_{a_1a_2}(D_i^{ }F_{ij})^{a_1}(D_k^{ }F_{kj})^{a_2}$
to be traded for combinations of lower-point operators via the field redefinition
$R_{\scalind{1}}^{ } \to R_{\scalind{1}}^{ } + \delta R_{\scalind{1}}^{ }$,
with $\delta R_{\scalind{1}}^{ }$ chosen to cancel the EOM-proportional operator.
This elimination can be performed at the expense of
shifting the Wilson coefficient tensors $O^{(n,k)}$ of the remaining operators.
In this work, however, we do not perform such redefinitions,
and all results are presented in terms of the operator basis defined above.
Possible extensions of the software to include field redefinitions are
left for future updates.
Indeed, it is in principle possible to construct
a non-redundant operator basis, analogous to
eqs.~\eqref{eq:dim_5_operators} and~\eqref{eq:dim_6_operators},
where each Wilson coefficient tensor $O^{(n,k)}$ is a linear combination
of the tensors in the redundant basis, including contributions
from the eliminated operators.
This program has already been carried out for
zero-temperature EFTs in~\cite{Fonseca:2025zjb};
and recently extended
to finite-temperature
using functional~\cite{Fuentes-Martin:2026bhr} and
Hilbert series~\cite{Chakrabortty:2026swu} methods.

To automate the matching,
all tensors $O$
in eqs.~\eqref{eq:dim_5_operators} and~\eqref{eq:dim_6_operators}
can be expressed in
terms of the tensors
$g^{abc}$, $g^{a}_{IJ}$, \dots, $Y_{\scalind{}IJ}$,
which are already implemented in \dralgo{}
(cf.\ tab.~\ref{tab:coupling_tensors}).
A typical dimension-six tensor arising from
higher-order derivative terms of scalar two-point functions reads
\begin{align}
\label{eq:matching_examples}
    O^{(6,15)}_{\scalind{1}\scalind{2}} &=
      \frac{(3\xi+5)}{3}g^{a_1}_{\scalind{1}j_1} g^{a_1}_{\scalind{2}j_1} \mathcal{I}^b_3
    - \frac{1}{20}\lambda_{\scalind{1}j_1j_2}\lambda_{\scalind{2}j_1j_2}\mathcal{I}^b_4
    \nn &
    - \frac{1}{6}\bigl(Y_{\scalind{1}}^{J_1J_2}Y_{\scalind{2}J_1J_2}+Y_{\scalind{2}}^{J_1J_2}Y_{\scalind{1}J_1J_2}\bigr)\mathcal{I}^f_3
    \,,
\end{align}
with the thermal one-loop master integrals being
\begin{align}
\label{eq:Ib:If}
\mathcal{I}^{b,\alpha}_{s} &\equiv
    \sumint{P}'\frac{(p_0^2)^{\alpha}}{[P^2]^s}=\left(\frac{\Lambda^2e^\gammaE}{4\pi}\right)^{\epsilon}2T\frac{[2\pi T]^{d-2s+2\alpha}}{(4\pi)^{d/2}}\frac{\Gamma(s-\frac{d}{2})}{\Gamma(s)}\zeta_{2s-2\alpha-d}
    \,,\nn[2mm]
    \mathcal{I}^{f,\alpha}_{s} &\equiv
    \sumint{\{P\}}'\frac{(p_0^2)^{\alpha}}{[P^2]^s}=(2^{2s-2\alpha-d}-1)\,\mathcal{I}^{b,\alpha}_s
    \,,
\end{align}
where
a primed integral denotes the absence of a zero mode,
$\{P\}$ indicates fermionic momenta,
$\Lambda$ is the $\overline{\text{MS}}$ renormalization scale,
and $\mathcal{I}^{b/f}_{s}= \mathcal{I}^{b/f,0}_{s}$
with $d = 3-2\epsilon$.
Here, $\gammaE$ denotes the Euler-Mascheroni constant and
$\zeta_s = \zeta(s)$ the Riemann $\zeta$-function.
The computation is performed in
the background-field $R_\xi$ gauge~\cite{Abbott:1980hw,Martin:2018emo},
so the matching relations may contain explicit dependence on
the gauge-fixing parameter $\xi$, as illustrated
in eq.~\eqref{eq:matching_examples};
further details are provided in~\ref{gauge_dependence}.

\subsection{New functionalities for higher-dimensional operators matching
  in \dralgo{}}
\label{sec:new_functions}

In the updated version
\dralgo{}
\DRalgoVersion{},
the matching relations such as eq.~\eqref{eq:matching_examples} are implemented
through the functions
\makeatletter
\renewcommand\thelstlisting{Fun.\@arabic\c@lstlisting}
\makeatother
\setcounter{lstlisting}{0}
\begin{lstlisting}[
		label={lst:Dimension56Matching},
		caption={~},
		language=Mathematica,
		escapeinside={(*@}{@*)}
	]
Dimension5Matching[<TensorList>,<TensorNumberList>,<WilsonCoefficients>,<Dimension>]
Dimension6Matching[<TensorList>,<TensorNumberList>,<WilsonCoefficients>,<Dimension>]
\end{lstlisting}
Both functions depend on the following four arguments:
\begin{itemize}
    \item
    {\tt <TensorList>} [{\it List}\,]:\\
    list of group tensors to match,
    ordered consistently with {\tt <TensorNumberList>};
    
    \item
    {\tt <TensorNumberList>} [{\it List of Integers}\,]:\\
    indices of the operators to match, according to the bases in
    eqs.~\eqref{eq:dim_5_operators} and~\eqref{eq:dim_6_operators};
    
    \item
    {\tt <WilsonCoefficients>} [{\it List of Symbols}\,]:\\
    list of Wilson coefficients to be determined;
    
    \item
    {\tt <Dimension>} [{\it Integer or Symbol}\,]:\\
    number of spatial dimensions.
    Can be kept symbolic or set to a fixed value.
    In the presence of fermions, especially for parity-violating interactions,
    it is advisable to set $d=3$.
\end{itemize}
In summary, the matching of Wilson coefficients requires
the user to construct the group tensor structures in terms of the desired Wilson coefficients,
specify which operators from the basis are to be matched,
provide the list of variables to solve for,
and set the number of spatial dimensions.

It is also possible to inspect directly the structure of the operators $O$,
without providing explicit tensors for matching,
using the functions
\begin{lstlisting}[
    label={lst:ODIM56},
    language=Mathematica, caption={~},
    escapeinside={(*@}{@*)}]
ODIM5[<OperatorNumber>,<Dimension>]
ODIM6[<OperatorNumber>,<Dimension>]
\end{lstlisting}
which depend on two arguments:
\begin{itemize}
    \item {\tt <OperatorNumber>} [{\it Integer}\,]:
      index of the operator, referring to the bases in eqs.~\eqref{eq:dim_5_operators} and~\eqref{eq:dim_6_operators};
    \item {\tt <Dimension>} [{\it Integer or Symbol}\,]:
      number of spatial dimensions, fixed or symbolic.
\end{itemize}

All results are expressed in terms of
the hard thermal one-loop integrals
\begin{align}
    {\tt Zb[s,0]} &\equiv \mathcal{I}_{s}^{b,\,0} \,, &
    {\tt Zf[s,0]} &\equiv \mathcal{I}_{s}^{f,\,0} \,,
\end{align}
which can be evaluated through the function
\begin{lstlisting}[
	language=Mathematica,
	caption={~},
	escapeinside={(*@}{@*)}
	]
HardThermal1LoopInt[<statistics>,<s>,<alpha>,<Dimension>]
\end{lstlisting}
which is defined for the
bosonic ({\tt "B"}) and
fermionic ({\tt "F"}) cases,
with the number of spatial dimensions ${\tt d} = d$ kept explicit,
as
\begin{align} 
    \texttt{HardThermal1LoopInt["B",$s$,$\alpha$,d]}
    &\equiv \mathcal{I}_{s}^{{\tt B},\alpha}
    = \mathcal{I}_{s}^{b,\alpha} \,, \\
    \texttt{HardThermal1LoopInt["F",$s$,$\alpha$,d]}
    &\equiv \mathcal{I}_{s}^{{\tt F},\alpha}
    = \mathcal{I}_{s}^{f,\alpha} \,.
\end{align}
{\tt HardThermal1LoopInt}
depends on four arguments:
\begin{itemize}
  \item {\tt <statistics>} [{\it String}\,]:
    {\tt "B"} (bosonic) or {\tt "F"} (fermionic);
  \item {\tt <s>} [{\it Integer}\,]:
    power of the propagator in the denominator;
  \item {\tt <alpha>} [{\it Integer}\,]:
    power of the Matsubara frequency in the numerator;
  \item {\tt <Dimension>} [{\it Integer or Symbol}\,]:
    number of spatial dimensions, fixed or symbolic.
\end{itemize}

To facilitate the construction of operator tensors,
we also introduce two additional helper functions,
\begin{lstlisting}[
  label={lst:ContractSymmetrize},
  language=Mathematica,
  caption={~},
  escapeinside={(*@}{@*)}
  ]
Contract[<TensorList>,<Indices>]
SymmetrizeTensor[<TensorList>,<Indices>]
\end{lstlisting}
which depend on two arguments:
\begin{itemize}
  \item {\tt <TensorList>} [{\it List}\,]:
    the array of tensors to be contracted or symmetrized;
  \item {\tt <Indices>} [{\it List}\,]:
    index pairs to contract, or indices over which to symmetrize.
\end{itemize}

A more pedagogical introduction to the implementation of these functions
is provided in the next section through explicit applications to specific models.
\newcounter{funlistingcount}
\setcounter{funlistingcount}{\value{lstlisting}}
\makeatletter
\renewcommand\thelstlisting{L.\the\numexpr\value{lstlisting}-\value{funlistingcount}\relax}
\makeatother

\section{Application to specific models}
\label{sec:application_to_models}

In this section, we demonstrate how to use the functions introduced 
in sec.~\ref{sec:new_functions}
to perform the matching of higher-dimensional operators in the
scalar-Yukawa model, QCD, and the full Standard Model.

\subsection{Scalar-Yukawa model}
\label{sec:firstExample}

To illustrate the use of the functions
{\tt Dimension5Matching} and
{\tt Dimension6Matching}
of~\ref{lst:Dimension56Matching},
we consider the matching of the complete set of
dimension-five and -six operators in a simple example,
namely the scalar-Yukawa model,
\begin{equation}
    \mathcal{L}^{(4d)}=
				\frac{1}{2}\phi_{4}(-\Box+m^2)\phi_{4}
			+ \overline{\psi}(i\slashed{\partial}-M)\psi
			+ \kappa\phi_4^3+\lambda\phi_4^4
			+ y\,\overline{\psi}\psi\phi_4
			\;.
\end{equation}
This model, sometimes referred to as the singlet fermionic dark matter model,
has been extensively studied in the context of
dark matter phenomenology~\cite{%
	Kim:2008pp,Baek:2011aa,Baek:2012uj,Esch:2013rta,Buchmueller:2013dya,Klasen:2013ypa,
	Esch:2014jpa,Freitas:2015hsa,Baek:2015lna,Dupuis:2016fda,Albert:2016osu,Bell:2016ekl},
and its phase transition structure has been
investigated in~\cite{Fairbairn:2013uta,Li:2014wia,Beniwal:2018hyi}.
The corresponding three-dimensional EFT
truncated at dimension four was investigated in~\cite{Gould:2023jbz},
while the matching of the dimension-five and -six operators was performed in~\cite{Chala:2024xll}.
The corresponding operator basis
of~\cite{Chala:2024xll},%
\footnote{%
    The Wilson coefficients used here differ in normalization from those in~\cite{Chala:2024xll}.
    The normalization factors are chosen such that the coefficients
    become canonically normalized when mapped onto the basis in
    eq.~\eqref{eq:dim_5_operators}.
} is given by
\begin{align}
    \mathcal{L}^{(3d)} &=
      \frac{1}{2}\phi(-\Box+m_3^2)\phi
      + \kappa_3^{ }\phi^3
      + \lambda_3^{ }\phi^4
      + \mathcal{L}^{(3d)}_{5}
      + \mathcal{L}^{(3d)}_{6}
    \,,
\end{align}
with
the higher-dimensional Lagrangians defined as
\begin{align}
\mathcal{L}^{(3d)}_{5}&=
		\frac{1}{5!}\alpha_{5,1}\phi^5
	+ \frac{1}{2!}\beta_{5,1}\phi^2\partial^2\phi
	\,,\nn
\mathcal{L}^{(3d)}_{6}&=
		\frac{1}{6!}\alpha_{6,1}\phi^6
	+ \frac{1}{2!}\beta_{6,1}(\partial^2\phi)(\partial^2\phi)
	- \frac{1}{4!}\beta_{6,2}\phi^3 \partial^2\phi
	\,.
\end{align}
Since the derivative operators above do not directly match the basis forms of
eqs.~\eqref{eq:dim_5_operators} and~\eqref{eq:dim_6_operators},
they must first be brought into that form before matching.
This step is a prerequisite for applying the functions
{\tt Dimension5Matching} and
{\tt Dimension6Matching}.

In this case, it is sufficient to
apply integration-by-parts identities to the derivative interactions, yielding
only a modification 
of the dimension-six operators,
\begin{align}
  \mathcal{L}^{(3d)}_{6}&=
      \frac{1}{6!}\alpha_{6,1}\phi^6
    + \frac{1}{2!}\beta_{6,1}\phi \partial^4\phi
    + \frac{1}{2!2!}\beta_{6,2}\phi^2 (\partial_\mu\phi)^2
  \,.
\end{align}
The next step is to identify the tensors $O$
appearing in eqs.~\eqref{eq:dim_5_operators} and~\eqref{eq:dim_6_operators}.
For the scalar-Yukawa model at dimension-five level,
they correspond to
\begin{align}
  O^{(5,2)}_{\scalind{1}\scalind{2}\scalind{3}\scalind{4}\scalind{5}}&=\alpha_{5,1}
    \,,&
    O^{(5,5)}_{\scalind{1}\scalind{2}\scalind{3}}&=\beta_{5,1}
    \,,
\end{align}
where, since the scalar-Yukawa model contains a single real scalar field,
all scalar indices $\scalind{}$ take only one value.
The explicit construction of these tensors as \texttt{Mathematica} objects, written in terms of the Wilson coefficients, is left to the user.
In this context, {\tt GroupMath}~\cite{Fonseca:2020vke}
may provide a useful framework for implementing the tensors.
The construction of tensors in {\tt Mathematica} can be facilitated through
the function {\tt TensorProduct},
which computes the tensor product of its arguments.
Alternatively, it can be used directly to construct tensors
with the desired dimensionality, as illustrated in the following script for the dimension-five operators:%
\footnote{%
  The script requires \dralgo{} to be installed and loaded,
  together with the scalar-Yukawa model file
  \href{https://github.com/DR-algo/DRalgo/examples/ScalYukawa_HDO.m}{\texttt{ScalYukawa\_HDO.m}},
  which defines the group-theory input tensors
  ($g^{abc}$, $g^a_{IJ}$, \dots) used below.
}
\begin{lstlisting}[
	language=Mathematica,
	escapeinside={(*@}{@*)}
	]
OT[5,2]=(*@$\alpha$@*)[5,1] TensorProduct[{1},{1},{1},{1},{1}];
OT[5,5]=(*@$\beta$@*)[5,1] TensorProduct[{1},{1},{1}];
\end{lstlisting}
We can now employ the function
{\tt Dimension5Matching}.
The following script performs the matching of the dimension-five operators
and returns the corresponding Wilson coefficients,
\begin{lstlisting}[
		language=Mathematica,
		escapeinside={(*@}{@*)}
	]
TensorList5={OT[5,2],OT[5,5]};
NList5={2,5}; 
WC5 = {(*@$\alpha$@*)[5,1],(*@$\beta$@*)[5,1]}; 
sol5=Dimension5Matching[TensorList5,NList5,WC5,d][[1]];
Collect[sol5,{_Zb,_Zf},Factor]//TableForm
\end{lstlisting}
\begin{outputlisting}[language=Mathematica, escapeinside={(*@}{@*)}]
(*@$\alpha$@*)[5,1]-> 51840 (*@$T^{3/2}\,\kappa\,\lambda^2$@*) Zb[3,0] -155520 (*@$T^{3/2}\,\kappa^3\,\lambda$@*) Zb[4,0]+93312 (*@$T^{3/2}\,\kappa^5$@*) Zb[5,0]
(*@$\beta$@*)[5,1]-> -24 (*@$T^{1/2}\,\kappa\,\lambda$@*) Zb[3,0]+54 (*@$T^{1/2}\,\kappa^3$@*) Zb[4,0]
\end{outputlisting}

For the matching of dimension-six operators,
the operators $O$ correspond to
\begin{align}
    O^{(6,4)}_{\scalind{1}\scalind{2}\scalind{3}\scalind{4}}&=\beta_{6,2}
    \,,&
    O^{(6,9)}_{\scalind{1}\scalind{2}\scalind{3}\scalind{4}\scalind{5}\scalind{6}}&=\alpha_{6,1}
    \,,&
    O^{(6,15)}_{\scalind{1}\scalind{2}}&=\beta_{6,1}\,,
\end{align}
which are associated with the
$3$-, $6$-, and $2$-point operators,
and can be constructed using 
\begin{lstlisting}[
		language=Mathematica,
		escapeinside={(*@}{@*)}
	]
OT[6,4]=(*@$\beta$@*)[6,2] TensorProduct[{1},{1},{1},{1}];
OT[6,9]=(*@$\alpha$@*)[6,1] TensorProduct[{1},{1},{1},{1},{1},{1}];
OT[6,15]=(*@$\beta$@*)[6,1] TensorProduct[{1},{1}];
\end{lstlisting}
To find the values of
$\alpha_{6,1}$,
$\beta_{6,1}$, and
$\beta_{6,2}$,
we can implement the following script
to find the Wilson coefficients,
\begin{lstlisting}[
		language=Mathematica,
		escapeinside={(*@}{@*)}
	]
TensorList6={OT[6,4],OT[6,9],OT[6,15]};
NList6={4,9,15};
WC6 = {(*@$\alpha$@*)[6,1],(*@$\beta$@*)[6,1],(*@$\beta$@*)[6,2]};
sol6=Dimension6Matching[TensorList6,NList6,WC6,d][[1]];
Collect[sol6,{_Zb,_Zf},Factor]//TableForm
\end{lstlisting}
\begin{outputlisting}[language=Mathematica, escapeinside={(*@}{@*)}]
(*@$\alpha$@*)[6,1]-> 207360 (*@$T^2\,\lambda^3$@*) Zb[3,0]-2799360 (*@$T^2\,\kappa^2\,\lambda^2$@*) Zb[4,0]+5598720 (*@$T^2\,\kappa^4\,\lambda$@*) Zb[5,0]
         -2799360 (*@$T^2\,\kappa^6$@*) Zb[6,0]-480 (*@$T^2\,y^6$@*) Zf[3,0]
(*@$\beta$@*)[6,1]-> (*@$-\frac{9}{5}\,\kappa^2$@*) Zb[4,0]-(*@$\frac{2}{3}\,y^2$@*) Zf[3,0]
(*@$\beta$@*)[6,2]-> 192 (*@$T\,\lambda^2$@*) Zb[3,0]-2160 (*@$T\,\kappa^2\,\lambda$@*) Zb[4,0]+2592 (*@$T\,\kappa^4$@*) Zb[5,0]-(*@$\frac{40}{3}\,T\,y^4$@*) Zf[3,0]
\end{outputlisting}
The results for the scalar-Yukawa model agree with
those reported in~\cite{Chala:2024xll}.

The matching can also be performed on a single operator or
on a subset of the full operator basis using~\ref{lst:Dimension56Matching}.
For instance, to obtain only the expressions for
the dimension-five $\alpha_{5,1}$
(cf.~\ref{lst:single:dim5}),
or
dimension-six $\beta_{6,1}$ and $\beta_{6,2}$
(cf.~\ref{lst:single:dim6}), respectively,
one can use the following scripts
\begin{lstlisting}[
		label={lst:single:dim5},
		language=Mathematica,
		escapeinside={(*@}{@*)}
	]
sol=Dimension5Matching[{OT[5,2]},{2},{(*@$\alpha $@*)[5,1]},d][[1]];  
sol//Factor//TableForm
\end{lstlisting}
\begin{outputlisting}[language=Mathematica, escapeinside={(*@}{@*)}]
(*@$\alpha$@*)[5,1]-> 10368 (*@$T^{3/2}\,\kappa$@*) (5(*@$\lambda^2$@*) Zb[3,0]-15(*@$\kappa^2\,\lambda$@*) Zb[4,0]+9(*@$\kappa^4$@*) Zb[5,0]) 
\end{outputlisting}
\begin{lstlisting}[
	label={lst:single:dim6},
	language=Mathematica,
	escapeinside={(*@}{@*)}
	]
sol=Dimension6Matching[{OT[6,4],OT[6,15]},{4,15},{(*@$\beta$@*)[6,1],(*@$\beta$@*)[6,2]},d][[1]];
sol//Factor//TableForm
\end{lstlisting}
\begin{outputlisting}[
		language=Mathematica,
		escapeinside={(*@}{@*)}
	]
(*@$\beta$@*)[6,1]-> (*@$\frac{1}{15}$@*) (-27(*@$\kappa^2$@*) Zb[4,0]-10(*@$y^2$@*) Zf[3,0])
(*@$\beta$@*)[6,2]->  (*@$\frac{8}{3}\,T$@*) (72(*@$\lambda^2$@*) Zb[3,0]-810(*@$\kappa^2\,\lambda$@*) Zb[4,0]+972(*@$\kappa^4$@*) Zb[5,0]-5(*@$y^4$@*) Zf[3,0])
\end{outputlisting}
Furthermore, it is also possible to inspect the shape of
the tensors $O$ directly, without providing a corresponding tensor to match to,
using the functions
{\tt ODIM5}, and
{\tt ODIM6}
of~\ref{lst:ODIM56}.
For the scalar-Yukawa model,
the following script 
returns the structure of the operator $O^{(5,5)}$,
\begin{lstlisting}[
	language=Mathematica,
	escapeinside={(*@}{@*)}
]
ODIM5[5,d]//MatrixForm
\end{lstlisting}
\begin{outputlisting}[language=Mathematica, escapeinside={(*@}{@*)}]
((6 Sqrt[T] (-4 (*@$\kappa\,\lambda$@*) Zb[3,0]+9 (*@$\kappa^3$@*) Zb[4,0])))
\end{outputlisting}

The scalar-Yukawa model including
the dimension-five and -six operator matching
is available at
\verb|https://github.com/DR-algo/DRalgo/examples/ScalYukawa_HDO.m|.

%
\subsection{Dimension-six Lagrangian of hot QCD}
\label{sec:QCD}

As a more complex example,
we consider the dimension-six Lagrangian of hot QCD
with one quark flavor,
\begin{equation}
	\label{eq:QCD_lag}
    \mathcal{L}^{(4d)}_{\text{QCD}}=
				\frac{1}{4}F^a_{\mu\nu}F^a_{\mu\nu}
			+ \Bar{\psi}\slashed D\,\psi
    \,.
\end{equation}
The pure bosonic higher-dimensional operators
of dimensionally reduced hot Yang-Mills were studied in~\cite{Chapman:1994vk,Laine:2018lgj},
while the fermionic contributions
to the thermal effective parameters were computed in~\cite{Megias:2003ui}
using the Heat-kernel method~\cite{%
  Avramidi:1997jy,Megias:2002vr,Vassilevich:2003xt,Chakrabortty:2024wto}.

The Lagrangian~\eqref{eq:QCD_lag} is defined in \dralgo{}
through the following group and field content:
\begin{lstlisting}[
	language=Mathematica,
	escapeinside={(*@}{@*)}
	]
Group={"SU3"};
CouplingName={gs};
RepAdjoint={{1,1}};
RepScalar={};
RepFermionL={{{1,0}},"L"};
RepFermionR={{{1,0}},"R"};
RepFermion={RepFermionL,RepFermionR};
{gvvv,gvff,gvss,(*@$\lambda$@*)1,(*@$\lambda$@*)3,(*@$\lambda$@*)4,(*@$\mu$@*)ij,(*@$\mu$@*)IJ,(*@$\mu$@*)IJC,Ysff,YsffC}=
  AllocateTensors[Group,RepAdjoint,CouplingName,RepFermion,RepScalar];
\end{lstlisting}
Dimensional reduction of the model from
the hard to the soft scale is then performed via
the following \dralgo{} commands.
We restrict to integrating out only the
hard scale $\pi T$ and do not proceed to lower scales.
The resulting effective theory is the three-dimensional EFT of hot QCD,
which is referred to as electrostatic QCD (EQCD)~\cite{Braaten:1995cm,Braaten:1995jr},
\begin{lstlisting}[
	language=Mathematica,
	escapeinside={(*@}{@*)}
	]
ImportModelDRalgo[Group,gvvv,gvff,gvss,(*@$\lambda$@*)1,(*@$\lambda$@*)3,(*@$\lambda$@*)4,(*@$\mu$@*)ij,(*@$\mu$@*)IJ,(*@$\mu$@*)IJC,Ysff,YsffC,Verbose->False];
PerformDRhard[]
\end{lstlisting}
We now perform the higher-dimensional operator matching in \dralgo{}.
The first step is to specify the operator basis%
\footnote{%
    The double-trace operators in the last two lines
    do not appear in
    the ${\rm SU}(3)$ QCD literature at one loop.
    There,
    $\alpha_{A_0^2F^2,3}$ vanishes for $\Nc=3$
    but is non-zero for general $\Nc$
    (cf.\ ${\rm SU}(2)$~\cite{Megias:2003ui}),
    while $\alpha_{A_0^6,2}$ is evanescent in the presence of fermions.
    Nevertheless, we list all independent operators for generality.
    The coefficients $\alpha_i$ can be related to the coefficients
    $c_i$ used in~\cite{Laine:2018lgj}; this correspondence can
    be verified in appendix~\ref{appendix:QCD}.
}
which resembles the basis in eq.~\eqref{eq:dim_6_operators},
{\em viz.}
\begin{align}
\label{eq:off_shell_QCD_lag}
    \mathcal{L}_{6,\rmii{QCD}}^{(3d)}&=
    \tr\biggl\{
			\frac{1}{2!}\alpha_{D^2F^2}\,(D_iF_{ik})^2+\frac{1}{2!}\alpha_{D^4A_0^2}(D^2A_0)^2+\,\frac{1}{3!}\alpha_{F^3}\,  F_{ij}F_{jk}F_{ki}+\frac{1}{2!2!}\alpha_{A_0^2F^2,1}A_0^2F_{ij}^2
    \nn[1mm] &
    \hphantom{{}=\tr\biggl\{}
    +\frac{1}{2!2!}\alpha_{A_0^2F^2,2}A_0F_{ij}A_0F_{ij}
    +\frac{1}{2!2!}\alpha_{D^2A_0^4,1}A_0^2(D_iA_0)^2
    +\frac{1}{2!2!}\alpha_{D^2A_0^4,2}(A_0D_iA_0)^2
    \nn[1mm] &
    \phantom{{}=\tr\biggl\{}
    +\frac{1}{6!}\alpha_{A_0^6,1}\,A_0^6
    +\frac{1}{2!}\alpha_{D^2A_0^2F}(D_iF_{ji})((D_jA_0)A_0-A_0(D_jA_0))\biggr\}
    \nn &
    +\frac{1}{6!}\alpha_{A_0^6,2}\tr\left\{A_0^4\right\}\tr\left\{A_0^2\right\}
    +\frac{1}{2!2!}\alpha_{A_0^2F^2,3}\tr\left\{A_0^2\right\}\tr\left\{F_{ij}^2\right\}
    \nn &
		+ \frac{1}{2!2!}\alpha_{D^2A_0^2,3}\tr\left\{(D_iA_0)^2\right\}\tr\left\{A_0^2\right\}
    \,.
\end{align}
By 
The corresponding group tensors are
that enter eq.~\eqref{eq:dim_6_operators} are
\begin{align}
    O^{(6,1)}_{\suiii{1}\suiii{2}\suiii{3}}&=
    \alpha_{F^3}\,X_{\suiii{1}\suiii{2}\suiii{3}}
    \,,\nn
    O^{(6,3)}_{\suiii{1}\suiii{2}\suiii{3}\suiii{4}}&=
      \Bigl[
          \alpha_{A_0F^2,1}\,X_{\suiii{1}\suiii{2}\suiii{3}\suiii{4}}
        + \alpha_{A_0F^2,2}\,X_{\suiii{1}\suiii{2}\suiii{3}\suiii{4}}
        + \alpha_{A_0^2F^2,3}\,X_{\suiii{1}\suiii{2}}X_{\suiii{3}\suiii{4}}
      \Bigr]_{({\suiii{1}\suiii{2}),(\suiii{3}\suiii{4})}}
      \,,\nn
    O^{(6,8)}_{\suiii{1}\suiii{2}\suiii{3}\suiii{4}}&=
      \Bigl[
          \alpha_{D^2A_0^4,1}\,X_{\suiii{1}\suiii{2}\suiii{3}\suiii{4}}
        + \alpha_{D^2A_0^4,2}\,X_{\suiii{1}\suiii{2}\suiii{3}\suiii{4}}
        + \alpha_{D^2A_0^4,3}\,X_{\suiii{1}\suiii{2}}X_{\suiii{3}\suiii{4}}
      \Bigr]_{(\suiii{1}\suiii{2}),(\suiii{3}\suiii{4})}
      \,,\nn
    O^{(6,12)}_{\suiii{1}\suiii{2}\suiii{3}\suiii{4}\suiii{5}\suiii{6}}&=
      \Bigl[
          \alpha_{A_0^6,1}\,X_{\suiii{1}\suiii{2}\suiii{3}\suiii{4}\suiii{5}\suiii{6}}
        + \alpha_{A_0^6,2}\,X_{\suiii{1}\suiii{2}\suiii{3}\suiii{4}}\,X_{\suiii{5}\suiii{6}}
      \Bigr]_{(\suiii{1}\suiii{2}\suiii{3}\suiii{4}\suiii{5}\suiii{6})}
    \,,\nn[1mm]
    O^{(6,13)}_{\suiii{1}\suiii{2}}&=\,\alpha_{D^2F^2}\,X_{\suiii{1}\suiii{2}}
    \,,\nn[1mm]
    O^{(6,16)}_{\suiii{1}\suiii{2}\suiii{3}}&=\,\alpha_{D^2A_0^2F}\,X_{\suiii{1}\suiii{2}\suiii{3}}
    \,,\nn[1mm]
    O^{(6,17)}_{\suiii{1}\suiii{2}}&=\,\alpha_{D^2A_0^2}\,X_{\suiii{1}\suiii{2}}
    \,,
\end{align}
where we organized the group generators into the following tensors 
\begin{equation}
    X^{\suiii{1}...\suiii{n}}=\tr\bigl\{T^{\suiii{1}}\dots T^{\suiii{n}}\bigr\}
    \,,
\end{equation}
with $T^{\suiii{1}}_{\suiii{2}\suiii{3}}=-i\,f^{\suiii{1} \suiii{2} \suiii{3}} $
being
the generators in the adjoint representation and
having used the following abbreviation
\begin{align}
    \left[X^{...i...j...}\right]_{(i,j)}&\equiv \frac{1}{2}\left[X^{...i...j...}+X^{...j...i...}\right]
		\,,&
    \left[X^{...i...j...}\right]_{[i,j]}&\equiv \frac{1}{2}\left[X^{...i...j...}-X^{...j...i...}\right]
    \,,
\end{align}
to (anti)symmetrize the tensors.
Each tensor must satisfy specific symmetry properties
for the function {\tt Dimension6Matching} to be applied consistently.
See~\ref{sec:symmetry_prop} for a more detailed discussion.
We can construct the previous tensors as follows:
\begin{lstlisting}[language=Mathematica, escapeinside={(*@}{@*)}]
Tadj=-I gvvv/gs;
X2=Contract[Tadj,Tadj,{{2,6},{3,5}}];
X3=Contract[Tadj,Tadj,Tadj,{{2,9},{3,5},{6,8}}];
X4=Contract[Tadj,Tadj,Tadj,Tadj,{{2,12},{3,5},{6,8},{9,11}}];
X6=Contract[Tadj,Tadj,Tadj,Tadj,Tadj,Tadj,{{2,18},{3,5},{6,8},{9,11},{12,14},{15,17}}];

OT[6,1]=(*@$\alpha$@*)[F^3] X3;
OT[6,3]=SymmetrizeTensor[
    +(*@$\alpha$@*)[A^2F^2,1]X4
    +(*@$\alpha$@*)[A^2F^2,2]Transpose[X4,{1,3,2,4}]
    +(*@$\alpha$@*)[A^2F^2,3]TensorProduct[X2,X2],{{1,2},{3,4}}
  ];
OT[6,8]=SymmetrizeTensor[
    +(*@$\alpha$@*)[D^2A^4,1]X4
    +(*@$\alpha$@*)[D^2A^4,2]Transpose[X4,{1,3,2,4}]
    +(*@$\alpha$@*)[D^2A^4,3]TensorProduct[X2,X2],{{1,2},{3,4}}
  ];
OT[6,12]=SymmetrizeTensor[
    +(*@$\alpha$@*)[A^6,1]X6
    +(*@$\alpha$@*)[A^6,2]TensorProduct[X4,X2],{{1,2,3,4,5,6}}
  ];
OT[6,13]=(*@$\alpha$@*)[D^2F^2]X2;
OT[6,16]=(*@$\alpha$@*)[D^2 A^2 F]X3;
OT[6,17]=(*@$\alpha$@*)[D^4A^2]X2;
\end{lstlisting}
We constructed the adjoint generators using the tri-vector tensors {\tt gvvv}, and,
to facilitate the construction of the tensors,
we implemented the functions
{\tt Contract} and
{\tt SymmetrizeTensor}
previously defined in~\ref{lst:ContractSymmetrize}.

By executing the following script,
\dralgo{} returns the QCD matching relations,
up to dimension six, in terms of the Wilson coefficients $\alpha_i$,
\begin{lstlisting}[
  language=Mathematica,
  escapeinside={(*@}{@*)}
  ]
TensorList={OT[6,1],OT[6,3],OT[6,8],OT[6,12],OT[6,13],OT[6,16],OT[6,17]};
NList={1,3,8,12,13,16,17};
WCs=DeleteDuplicates[Cases[TensorList//Normal, (*@$\alpha$@*)[__], \[Infinity]]];
sol6=Dimension6Matching[TensorList,NList,WCs,3][[1]]; 
Collect[sol6,{_Zb,_Zf},Factor]//TableForm
\end{lstlisting}
\begin{outputlisting}[
		language=Mathematica,
		escapeinside={(*@}{@*)}
	]
(*@$\alpha$@*)[F^3] -> -(*@$\frac{2}{15}$@*)I gs^3 Sqrt[T] Zb[3, 0] + (*@$\frac{2}{45}$@*) I gs^3 Sqrt[T] Zf[3,0]
(*@$\alpha$@*)[A^2 F^2,1] -> -(*@$\frac{146}{15}$@*) gs^4 Zb[3, 0] + (*@$\frac{74}{405}$@*) gs^4 Zf[3,0]
(*@$\alpha$@*)[A^2 F^2,2] -> -(*@$\frac{74}{15}$@*) gs^4 Zb[3, 0] + (*@$\frac{46}{405}$@*) gs^4 Zf[3,0]
(*@$\alpha$@*)[A^2 F^2,3] -> 0
(*@$\alpha$@*)[A^4 D^2,1] -> -(*@$\frac{1}{3}$@*) gs^4 T (45-2xi+xi^2) Zb[3,0] + (*@$\frac{8}{81}$@*) gs^4 T Zf[3,0]
(*@$\alpha$@*)[A^4 D^2,2] -> (*@$\frac{1}{3}$@*) gs^4 T (-43-2xi+xi^2) Zb[3,0] + (*@$\frac{40}{81}$@*) gs^4 T Zf[3, 0]
(*@$\alpha$@*)[D^2A^4,3] -> 0
(*@$\alpha$@*)[A^6,1] -> 0
(*@$\alpha$@*)[A^6,2] -> 0
(*@$\alpha$@*)[D^2 F^2] -> -(*@$\frac{1}{60}$@*) gs^2 (-111+30xi+5xi^2) Zb[3, 0] - (*@$\frac{4}{45}$@*) gs^2 Zf[3, 0]
(*@$\alpha$@*)[A^2 D^2 F]-> -(*@$\frac{1}{30}$@*) I gs^3 Sqrt[T] (-89+10xi+5xi^2) Zb[3, 0] - (*@$\frac{7}{45}$@*) I gs^3 Sqrt[T] Zf[3, 0]
(*@$\alpha$@*)[A^2 D^4]-> -(*@$\frac{1}{60}$@*) gs^2 (-53-30xi+15xi^2) Zb[3,0] - (*@$\frac{2}{45}$@*) gs^2 Zf[3, 0]
\end{outputlisting}
The matching result contains explicit dependence on
the gauge-fixing parameter {\tt xi};
see \ref{gauge_dependence} for clarification.
Setting ${\tt xi}=1$ reproduces the results of~\cite{Laine:2018lgj,Megias:2003ui}.
As discussed in~\cite{Bernardo:2025vkz,Chala:2025aiz},
Wilson coefficients in a redundant operator basis may depend on the gauge parameter,
but this dependence cancels after the appropriate field redefinitions are applied.
The explicit {\tt xi} dependence is therefore retained
as a proxy to verify this cancellation
once the operator redundancy is removed.

As introduced
in~\ref{lst:single:dim5} and~\ref{lst:single:dim6},
the matching can be performed on a subset of group tensors:
\begin{lstlisting}[language=Mathematica, escapeinside={(*@}{@*)}]
sol=Dimension6Matching[{OT[6,1]},{1},{(*@$\alpha$@*)[F^3]},3][[1]];
Collect[sol,{_Zb,_Zf},Factor]//TableForm
\end{lstlisting}
\begin{outputlisting}[
	language=Mathematica,
	escapeinside={(*@}{@*)}
]
(*@$\alpha$@*)[F^3] -> -(*@$\frac{2}{15}$@*)I gs^3 Sqrt[T] Zb[3, 0] + (*@$\frac{2}{45}$@*) I gs^3 Sqrt[T] Zf[3,0]
\end{outputlisting}
To inspect the shape of the tensors $O$,
we can use~\ref{lst:ODIM56} to directly query
the tensors $O$ through
{\tt ODIM5} and
{\tt ODIM6}.
Here, we show the example of the dimension-six operators
$O^{(6,1)}$ and
$O^{(6,17)}$,
\begin{lstlisting}[
	language=Mathematica,
	escapeinside={(*@}{@*)}
	]
ODIM6[1,3]
ODIM6[17,3]
\end{lstlisting}
which returns
an $8\times 8\times 8$ totally antisymmetric tensor for $O^{(6,1)}$ and
an $8\times 8$ tensor proportional to the identity for $O^{(6,17)}$.

Since the operator basis~\eqref{eq:dim_6_operators} is redundant,
a subset of operators can be eliminated via field redefinitions,
yielding the physical (on-shell) basis
\begin{align}
\label{eq:on_shell_QCD_lag}
\mathcal{L}_{\text{6,phys}}^{(3d)}&=
		\tr\biggl\{
					\frac{1}{3!}\beta_{F^3}\,F_{ij}F_{jk}F_{ki}
				+	\frac{1}{2!2!}\beta_{A_0^2F^2,1}\,A_0^2F_{ij}^2
				+ \frac{1}{2!2!}\beta_{A_0^2F^2,2}\,A_0F_{ij}A_0F_{ij}
		\nn &\phantom{{}=\tr\biggl\{}
			+ \frac{1}{2!2!}\beta_{D^2A_0^4,1}(A_0D_iA_0)^2
			+ \beta_{A_0^6,1}\,A_0^6
			\biggr\}
		\nn &
		+ \frac{1}{6!}\beta_{A_0^6,2}\tr\left\{A_0^4\right\}\tr\left\{A_0^2\right\}
    + \frac{1}{2!2!}\beta_{A_0^2F^2,3}
			\tr\left\{A_0^2\right\}
			\tr\left\{F_{ij}^2\right\}
    \nn &
		+ \frac{1}{2!2!}\beta_{D^2A_0^4,2}\bigl(\tr\left\{A_0D_iA_0\right\}\bigr)^2
    \,,
\end{align}
together with 
the redefined Wilson coefficients $\beta_i$,
\begin{align}
  \label{eq:QCD_Wilson_coefficients}
    \beta_{F^3}&=-\frac{2i}{45}\Bigl(
          3\,\mathcal{I}^b_3
        - \Nf^{ }\,\mathcal{I}^f_3
      \Bigr)\gs^3\,T^{1/2}
		\,,&
    \beta_{A_0^2F^2,1}&=
    \frac{2}{405}\Bigl(
      - 1971\,\mathcal{I}^b_3
      + 23\,\Nf^{ }\,\mathcal{I}^f_3
      \Bigr)\gs^4
		\,,\nn
    \beta_{A_0^2F^2,2}&=-
    \frac{2}{405}\Bigl(
        999\,\mathcal{I}^b_3
      - 23\,\Nf^{ }\,\mathcal{I}^f_3
      \Bigr)\gs^4
		\,,&
    \beta_{D^2A_0^4,1}&=-\frac{2}{45}\Bigl(
        3\,\mathcal{I}^b_3
      - \Nf^{ }\,\mathcal{I}^f_3
      \Bigr)\gs^4\,T
		\,,\nn[1mm]
    \beta_{D^2A_0^4,2}&=0
		\,,&
    \beta_{A_0^6,1}&=0
		\,,\nn[1mm]
    \beta_{A_0^6,2}&=0
		\,,
\end{align}
where we replaced $\mathcal{I}^f_3 \to \Nf^{ }\,\mathcal{I}^f_3$
to extend the result to a generic number $\Nf$ of quarks. 
The relation between the $\beta_i$ and $\alpha_i$ coefficients
is given in eq.~\eqref{eq:QCD_FR}.
As shown in~\cite{Bernardo:2025vkz,Chala:2025aiz},
gauge independence in the physical basis is also manifest for the QCD Lagrangian.

The QCD model file, including the dimension-six operator matching and the field redefinitions,
is available at
\verb|https://github.com/DR-algo/DRalgo/examples/QCD_HDO.m|.

\subsubsection{Fermionic contributions to hard-to-soft matching}

As an application of the on-shell basis~\eqref{eq:on_shell_QCD_lag} derived above,
we discuss the role of fermionic higher-dimensional operators in
renormalizing results obtained from the soft-scale EQCD.

When matching the hard to the soft scale,
for the magnetostatic gauge coupling $\gM$~\cite{Gross:1980br},
bosonic modes contribute from all scales when renormalizing the result~\cite{Laine:2018lgj}.
Conversely, fermionic modes only contribute
when integrating out the hard scale due to their absence in
the soft theory.
The matching of
the renormalized electrostatic gauge coupling
$g_\rmii{ER}^2 = g^2/(\mathcal{Z}_\rmii{$B$}+Z_\rmii{$B$})$
is determined by
the hard and soft wavefunction renormalization factors
$\mathcal{Z}_\rmii{$B$}$ and $Z_\rmii{$B$}$, respectively.
As a result,
computing $g_\rmii{ER}$
to $\mathcal{O}(g^8)$ requires accounting for
fermionic contributions to the bosonic dimension-six operators,
which enter through one- and two-loop soft diagrams
contributing to $Z_\rmii{$B$}$.
The two-loop contributions exhibit
a UV divergence that cancels against an IR divergence
from the hard three-loop $g_\rmii{ER}$ computation.

While the bosonic part of this computation has
already been performed in~\cite{Laine:2018lgj},
the fermionic contributions have not yet been determined,
primarily because a fermionic three-loop computation of $g_\rmii{ER}$ is also required.
Such a computation
currently still lacks a complete set of contributing master sum-integrals
in the fermionic sector.
However, the purely bosonic sector
has already been extended to four loops~\cite{Navarrete:2024ruu}.
The purely bosonic hard part of $g_\rmii{ER}$ at three-loop level
is listed in~\cite{Ghisoiu:2013zoj}.
An analogous bottleneck exists for the Debye mass parameter $\mE$;
see~\cite{Moller:2012chx} for a list
of the required fermionic master integrals.

To use the results for the fermionic
contributions to the dimension-six operator basis in the soft theory
from eq.~\eqref{eq:QCD_Wilson_coefficients},
we reuse the two-loop soft results of~\cite{Laine:2018lgj}.
Due to the appearance
of new operators in the basis~\eqref{eq:off_shell_QCD_lag},
for general $\Nc \neq 3$,
we conduct the corresponding computation in
$\Nc = 3$ QCD,
as for this specific case relations for the trace of
the generators in the adjoint representation~\cite{Haber:2019sgz}
allow one to reduce the basis in~\eqref{eq:off_shell_QCD_lag} to
the one found in~\cite{Laine:2018lgj}.
To this end,
we compute
the soft/hard%
\footnote{%
    Following~\cite{Laine:2018lgj}, ``soft/hard'' refers to
    insertions of hard dimension-six operators within soft-scale diagrams.
} correction
$\delta Z_\rmii{$B$}$ with the addition of the fermionic
contributions to the dimension-six operators.
The resulting expression for $\delta Z_\rmii{$B$}$
at fixed $\Nc = 3$ is given by
\begin{align}
\label{eq:deltaZb}
  \delta Z_\rmii{$B$} &=
    \left(\frac{3 g_\rmii{ER}^2}{(4\pi)^2} \right)^3
    \left(
      \frac{1337\Nf}{9882} - \frac{1097}{1098} \right)
      \frac{61\zeta_3}{5 \epsilon} 
  \,.
\end{align}
We anticipate that the fermionic contributions to the dimension-six operators
will be relevant for a complete computation of $g_\rmii{ER}$ to $\mathcal{O}(g^8)$,
and provide an additional crosscheck for such a future computation.

\subsection{Standard Model higher-dimensional operators}
\label{sec:SM}

The Standard Model provides a test case for the automated 
higher-dimensional operator matching
with~\dralgo{},
both due to its large field content and its phenomenological relevance for
assessing the validity of high-temperature dimensional reduction~\cite{Chala:2025aiz,Kajantie:1997ky}.
It contains $12$ gauge fields
(%
$1$ associated with ${\rm U}(1)_\rmii{$Y$}$,
$3$ with ${\rm SU}(2)_\rmii{$L$}$, and
$8$ with ${\rm SU}(3)_\rmi{c}$),
and the explicit construction of the group tensors $O$ becomes increasingly involved
for such large field cases.
In~\ref{lst:SM_tensors_construction},
we detail a systematic procedure for their construction.

As an illustration of the capabilities of \dralgo{},
we perform the dimensional reduction of the Standard Model up to
dimension-six operators.
The electroweak sector computation has previously been carried out in~\cite{Chala:2025aiz,Kajantie:1997ky}.
Here we extend the analysis by determining operators mixing the strong and electroweak sectors,
partially computed in~\cite{Moore:1995jv}.
We organize the operator basis,
\begin{align}
    \label{eq:sm_dim6_mainbody}
    \mathcal{L}^{(3d)}_{6,\rmii{SM}} &=
      \sum_{i=1}^{N_\rmii{SM}}\mathcal{L}^{(6,i)}_\rmii{SM}
    \,,
\end{align}
with $\mathcal{L}^{(6,i)}_\rmii{SM}$ defined in
eq.~\eqref{eq:dim6_SM} and
$N_\rmii{SM} = 20$ being the total number of independent operators.

As a concrete example,
we focus on the following set of parity-violating operators:
\begin{align}
    \mathcal{L}^{(6,18)}_\rmii{SM} &=
    	i\alpha_{B_0 B^2 D}B_0 \widetilde{B}_k (D_l B_{lk})
    +	i\alpha_{B_0 W^2 D}B_0 \widetilde{W}^{\suii{1}}_k (D_l W^{\suii{1}}_{lk})
		+ i\alpha_{W_0 B\,W\,D,1}W^{\suii{1}}_0 \widetilde{B}_k (D_l W^{\suii{1}}_{lk})
		\nn[1mm] &
		+ i\alpha_{W_0 B\,W\,D,2}W^{\suii{1}}_0 \widetilde{W}^{\suii{1}}_k (D_l B_{lk})
		+ i\alpha_{W_0 W^2\,D}\epsilon^{\suii{1}\suii{2}\suii{3}}W^{\suii{1}}_0\widetilde{W}^{\suii{2}}_k (D_l W^{\suii{3}}_{lk})
		\nn[1mm] &
		+ i\alpha_{B_0 G^2 D}B_0 \widetilde{G}^{\suiii{1}}_k (D_l G^{\suiii{1}}_{lk})
		+ i\alpha_{G_0 B\,G\,D,1}G^{\suiii{1}}_0 \widetilde{B}_k (D_l G^{\suiii{1}}_{lk})
		\nn[1mm] &
		+ i\alpha_{G_0 B\,G\,D,2}G^{\suiii{1}}_0 \widetilde{G}^{\suiii{1}}_k (D_l B_{lk})
		+ i\alpha_{G_0 G^2 D,1}f^{\suiii{1}\suiii{2}\suiii{3}}G^{\suiii{1}}_0\widetilde{G}^{\suiii{2}}_k \bigl(D_l G^{\suiii{3}}_{lk}\bigr)
		\nn[1mm] &
		+ i\alpha_{G_0 G^2 D,2}d^{\suiii{1}\suiii{2}\suiii{3}}G^{\suiii{1}}_0\widetilde{G}^{\suiii{2}}_k \bigl(D_l G^{\suiii{3}}_{lk}\bigr)
		\,,
\end{align}
where
$d^{\suiii{1}\suiii{2}\suiii{3}}$ is the totally symmetric tensor of ${\rm SU}(3)_\rmi{c}$.
Operators of this type can arise when the underlying $4d$ theory contains sources of parity violation.
As pointed out in~\cite{Kajantie:1997ky},
they may also be generated in the presence of
a finite chemical potential~\cite{Bochkarev:1989kp,Gynther:2003za};
such scenarios are, however, not considered in the present work.

When loading the Standard Model example file
\href{https://github.com/DR-algo/DRalgo/examples/sm_HDO.m}{\texttt{sm\_HDO.m}} from the \dralgo{} examples folder,
the gauge fields are ordered as
$G^{\suiii{i} {}_{=1,\dots,8}}$,
$W^{\suii{i} {}_{=1,\dots,3}}$,
$B$.
This ordering must be taken into account
when constructing the operator tensors.
For instance, the tensor $O^{(6,18)}_{a_1a_2a_3}$ is
constructed as follows:

\begin{lstlisting}[label={lst:SM_tensors_construction},language=Mathematica, escapeinside={(*@}{@*)}]
(* identity on SU(2) gauge fields *)
idW = SparseArray[{{i_, i_} /; 9 <= i <= 11 -> 1}, {12, 12}];

(* identity on SU(3) gauge fields *)
idG = SparseArray[{{i_, i_} /; 1 <= i <= 8 -> 1}, {12, 12}];

(* identity on U(1) gauge field *)
idB = SparseArray[{{i_, i_} /; 12 <= i <= 12 -> 1}, {12, 12}];

(* projector on U(1) gauge field *)
ProjB = SparseArray[{{i_} /; 12 <= i <= 12 -> 1}, {12}];

(* SU(2)L group structure constant *)
fW=Coefficient[gvvv,gw];

(* SU(3)C group structure constant *)
fG=Coefficient[gvvv,gs];

(* Gell-Mann Matrices *)
lambda[1] = {{0, 1, 0}, {1, 0, 0}, {0, 0, 0}};
lambda[2] = {{0, -I, 0}, {I, 0, 0}, {0, 0, 0}};
lambda[3] = {{1, 0, 0}, {0, -1, 0}, {0, 0, 0}};
lambda[4] = {{0, 0, 1}, {0, 0, 0}, {1, 0, 0}};
lambda[5] = {{0, 0, -I}, {0, 0, 0}, {I, 0, 0}};
lambda[6] = {{0, 0, 0}, {0, 0, 1}, {0, 1, 0}};
lambda[7] = {{0, 0, 0}, {0, 0, -I}, {0, I, 0}};
lambda[8] = (1/Sqrt[3]) {{1, 0, 0}, {0, 1, 0}, {0, 0, -2}};

(* d^{abc} Tensor*)
dG = ConstantArray[0, {12, 12, 12}];
dG[[1 ;; 8, 1 ;; 8, 1 ;; 8]] = Table[
  (1/4) Tr[(lambda[a] . lambda[b] + lambda[b] . lambda[a]) . lambda[c]],
  {a, 8}, {b, 8}, {c, 8}
];

OT[6,18]= (
    + I (*@$\alpha$@*)[B0 B^2 D]*TensorProduct[idB,ProjB]
    + I (*@$\alpha$@*)[B0 W^2 D]*TensorProduct[ProjB,idW]
    + I (*@$\alpha$@*)[W0 B W D,1]*Transpose[TensorProduct[ProjB,idW],{2,1,3}]
    + I (*@$\alpha$@*)[W0 B W D,2]*TensorProduct[idW,ProjB]
    + I (*@$\alpha$@*)[W0 W^2 D]*fW
    + I (*@$\alpha$@*)[B0 G^2 D]*TensorProduct[ProjB,idG]
    + I (*@$\alpha$@*)[G0 B G D,1]*Transpose[TensorProduct[ProjB,idG],{2,1,3}]
    + I (*@$\alpha$@*)[G0 B G D,2]*TensorProduct[idG,ProjB]
    + I (*@$\alpha$@*)[G0 G^2 D,1]*fG
    + I (*@$\alpha$@*)[G0 G^2 D,2]*dG
    );	

WCs = Cases[OT[6,18]//Normal, (*@$\alpha$@*)[__], \[Infinity]];
\end{lstlisting}
The matching of
these parity-violating Wilson coefficients can be performed
via
\begin{lstlisting}[
	language=Mathematica,
	escapeinside={(*@}{@*)}
]
sol=Dimension6Matching[{OT[6,18]},{18},3][[1]];
Collect[sol,{_Zb,_Zf},Factor]//TableForm
\end{lstlisting}
\begin{outputlisting}[
    label={lst:SM_dim6_parity_violating_WCs},
    language=Mathematica,
    escapeinside={(*@}{@*)}
    ]
 (*@$\alpha$@*)[B0 B^2 D] -> -(1/8) gY^3 Sqrt[T] (3 Yd^3 + Ye^3 - 2 Yl^3 - 6 Yq^3 + 3 Yu^3) Zf[3,0]
 (*@$\alpha$@*)[B0 W^2 D] -> 1/4 gw^2 gY Sqrt[T] (Yl + 3 Yq) Zf[3, 0]
 (*@$\alpha$@*)[W0 B W D,1] -> 1/4 gw^2 gY Sqrt[T] (Yl + 3 Yq) Zf[3, 0]
 (*@$\alpha$@*)[W0 B W D,2] -> 1/4 gw^2 gY Sqrt[T] (Yl + 3 Yq) Zf[3, 0]
 (*@$\alpha$@*)[W0 W^2 D] -> 0
 (*@$\alpha$@*)[B0 G^2 D] -> -(1/4) I gs^2 gY Sqrt[T] (Yd - 2 Yq + Yu) Zf[3, 0]
 (*@$\alpha$@*)[G0 B G D,1] -> -(1/4) gs^2 gY Sqrt[T] (Yd - 2 Yq + Yu) Zf[3, 0]
 (*@$\alpha$@*)[G0 B G D,2] -> 1/4 gw^2 gY Sqrt[T] (Yl + 3 Yq) Zf[3, 0]
 (*@$\alpha$@*)[G0 G^2 D,1] -> 0
 (*@$\alpha$@*)[G0 G^2 D,2] -> 0
\end{outputlisting}
All these Wilson coefficients vanish for
the physical values of the hypercharges of the various fields,
namely
$Y_q = \frac{1}{3}$,
$Y_u = \frac{4}{3}$,
$Y_d = -\frac{2}{3}$,
$Y_\ell = -1$, and
$Y_e = -2$.
In particular, the hypercharge combinations
appearing in~\ref{lst:SM_dim6_parity_violating_WCs}
are the same as those entering
SM anomaly cancellation~\cite{Bouchiat:1972iq,Gross:1972pv}.
Hence, these operators do not appear in
the SM literature~\cite{Kajantie:1997ky},
although they are generated in a general computation.

The Standard Model including
the dimension-five and -six operator matching
is available at
\verb|https://github.com/DR-algo/DRalgo/examples/sm_HDO.m|.

\renewcommand{\thesubsubsection}{Technical aspect~\arabic{subsubsection}}
\section{Miscellaneous technical aspects}
\label{sec:details}

\subsubsection{Spacetime dimensions}
\label{sec:spacetime_dimensions}

The dimension-five and -six matching can be performed in a generic number $d$ of
spatial dimensions in the purely bosonic case.
This is controlled by the {\tt <Dimension>} argument
of the functions~\ref{lst:Dimension56Matching} and~\ref{lst:ODIM56}.
The following script shows the pure-gluonic sector of QCD,
where the one-loop matching relations depend explicitly on $d = {\tt d}$:
\begin{lstlisting}[
	language=Mathematica,
	escapeinside={(*@}{@*)}
]
sol=Dimension6Matching[TensorList,NList,WC,d][[1]];
Collect[sol,{_Zb,_Zf},Factor]//TableForm
\end{lstlisting}
\begin{outputlisting}[language=Mathematica, escapeinside={(*@}{@*)}]
(*@$\alpha$@*)[F^3] -> (*@$\frac{1}{15}$@*)(1-d)I gs^3 Sqrt[T] Zb[3, 0]
(*@$\alpha$@*)[A^2 F^2,1] -> -(*@$\frac{1}{15}$@*) (419-103 d+4d^2) gs^4 Zb[3, 0]
(*@$\alpha$@*)[A^2 F^2,2] -> -(*@$\frac{1}{15}$@*) (206-47 d+d^2)gs^4 Zb[3, 0]
(*@$\alpha$@*)[A^2 F^2,3] -> 0
(*@$\alpha$@*)[D^2 A^4,1] -> (*@$\frac{1}{15}$@*) (-303+32 d-20 d^2+6 d^3-530 xi+300 d xi-40 d^2 xi-5 xi^2) gs^4 T Zb[3,0]
(*@$\alpha$@*)[D^2 A^4,2] -> (*@$\frac{1}{15}$@*)  (-197+18 d-20 d^2+4 d^3-370 xi+180 d xi-20 d^2 xi+5 xi^2) gs^4 T Zb[3,0]
(*@$\alpha$@*)[D^2 A^4,3] -> 0
(*@$\alpha$@*)[A^6,1] -> 8(-5+d)(-3+d)(-1+d)^2 gs^6 T^2 Zb[3, 0]
(*@$\alpha$@*)[A^6,2] -> 0
(*@$\alpha$@*)[D^2 F^2] -> -(*@$\frac{1}{60}$@*) gs^2 (-117+2 d+30 xi+5 xi^2) Zb[3, 0]
(*@$\alpha$@*)[D^2 A^2 F] -> (*@$\frac{1}{30}$@*) (71+3 d+d^2+20 xi-10 d xi-5 xi^2) I gs^3 Sqrt[T]  Zb[3, 0]
(*@$\alpha$@*)[A^2 D^4] -> (*@$\frac{1}{60}$@*) gs^2 (2+11 d +2 d^2+120 xi -30 d xi -30 xi^2+5 d xi^2) Zb[3,0]
\end{outputlisting}
In the presence of fermionic fields,
the functions 
{\tt Dimension5Matching},
{\tt Dimension6Matching}
of~\ref{lst:Dimension56Matching}
are not guaranteed to return valid solutions for
the Wilson coefficients at $d\neq 3$, and results
obtained for generic dimensions may be incorrect.
The fermionic sector relies on the Chisholm identities
in three dimensions
for the Weyl matrices~\cite{Dreiner:2008tw},
\begin{align}
    \overline{\sigma}^{\mu}\sigma^{\nu}\overline{\sigma}^{\rho}
    &=
      \delta^{\mu\nu}\overline{\sigma}^\rho
    - \delta^{\mu\rho}\overline{\sigma}^\nu
    + \delta^{\nu\rho}\overline{\sigma}^{\mu}
    + \epsilon^{\mu\nu\rho\kappa}\overline{\sigma}^{\kappa}
    \,,\nn
    {\sigma}^{\mu}\overline{\sigma}^{\nu}{\sigma}^{\rho}
    &=
      \delta^{\mu\nu}{\sigma}^\rho
    - \delta^{\mu\rho}{\sigma}^\nu
    + \delta^{\nu\rho}{\sigma}^{\mu}
    - \epsilon^{\mu\nu\rho\kappa}{\sigma}^{\kappa}
    \,,
\end{align}
which are structurally identical in Euclidean and Minkowski spacetime.
These identities involve the four-dimensional
Levi-Civita tensor $\epsilon^{\mu\nu\rho\kappa}$,
which has no natural generalization
to an arbitrary number of spacetime dimensions.
This is in direct analogy with the problem of
defining $\gamma_5$ in dimensional
regularization~\cite{Fuentes-Martin:2025meq}.
One strategy is to replace Weyl spinors by
four-dimensional Dirac $\gamma$-matrices~\cite{Fuentes-Martin:2026bhr}, but
this reintroduces the $\gamma_5$ problem for chiral fermions,
which must then be handled within
the standard schemes~\cite{tHooft:1972tcz,Breitenlohner:1975hg}.

In \dralgo{}, we instead follow the standard strategy
of dimensional reduction,
where the fermionic Lorentz and spin algebra remains four-dimensional
while only the momentum integrals are analytically continued to
$d=3-2\epsilon$ spatial dimensions.

\subsubsection{Matching a single operator}

It is possible to match a single group tensor $O$,
or any subset of them, provided that all
independent group invariants are
included in the basis definition.
For instance, in the QCD script, if one omits one of the operators
$\tr\{A_0^2F_{ij}^2\}$ or
$\tr\{(A_0F_{ij})^2\}$, as in the following example,
no solution is found for $\alpha_{A_0^2F^2,1}$.
\begin{lstlisting}[
	language=Mathematica,
	escapeinside={(*@}{@*)}
]
O[6,3]= 4(*@$\alpha$@*)[A^2F^2,1] SymmetrizeTensor[X4,{{1,2},{3,4}}];
Dimension6Matching[{O[6,3]},{3},{(*@$\alpha$@*)[A^2F^2,1]},3]
\end{lstlisting}

\subsubsection{Symmetry properties of group tensors}
\label{sec:symmetry_prop}

Group tensors must be constructed with well-defined symmetry properties
to avoid redundancy.
For instance, for the operator
\begin{equation}
    \mathcal{L}_6^{(3d)} \supset \frac{1}{2!}O^{(6,15)}_{\scalind{1}\scalind{2}}(D_iD_i\phi_{\scalind{1}})(D_jD_j\phi_{\scalind{2}})
		\,,
\end{equation}
the antisymmetric component of $O^{(6,15)}_{\scalind{1}\scalind{2}}$ does not contribute.
We therefore require
the dimension-five tensors to satisfy the symmetry properties,
\begin{align}
    O^{(5,1)}_{\scalind{1} a_1a_2}& \text{ symmetric in } (a_1,a_2)
    \,,&
    O^{(5,5)}_{\scalind{1}\scalind{2}\scalind{3}}& \text{ symmetric in } (\scalind{2},\scalind{3})
    \,,\nn
    O^{(5,2)}_{\scalind{1}\scalind{2}\scalind{3}\scalind{4}\scalind{5}}& \text{ symmetric in  } (\scalind{1},\scalind{2},\scalind{3},\scalind{4},\scalind{5})
    \,,&
    O^{(5,6)}_{\scalind{1}a_2a_3}& \text{ symmetric in } (a_2,a_3)
    \,,\nn
    O^{(5,3)}_{\scalind{1}\scalind{2}\scalind{3}a_4a_5}& \text{ symmetric in } (\scalind{1},\scalind{2},\scalind{3}) \text{ and } (a_4,a_5)
    \,,&
    O^{(5,7)}_{\scalind{1}a_2a_3}& \text{ no symmetry required}
    \,,\nn
    O^{(5,4)}_{\scalind{1}a_2a_3a_4a_5}& \text{ symmetric in } (a_2,a_3,a_4,a_5)
    \,,&
    O^{(5,8)}_{\scalind{1}a_2a_3}& \text{ no symmetry required}
    \,,
\end{align}
whereas the dimension-six tensors must satisfy
\begin{align}
    O^{(6,1)}_{a_1a_2a_3}& \text{ antisymmetric in } (a_1,a_2,a_3)
    \,,&
    O^{(6,13)}_{a_1a_2}& \text{ symmetric in } (a_1,a_2)
    \,,\nn
    O^{(6,2)}_{\scalind{1}\scalind{2}a_3a_4}& \text{ symmetric in } (\scalind{1},\scalind{2}) \text{ and } (a_3,a_4)
    \,,&
    O^{(6,14)}_{\scalind{1}\scalind{2}a_3}& \text{ antisymmetric in } (\scalind{1},\scalind{2})
    \,,\nn
    O^{(6,3)}_{a_1a_2a_3a_4}& \text{ symmetric in } (a_1,a_2) \text{ and } (a_3,a_4)
    \,,&
    O^{(6,15)}_{\scalind{1}\scalind{2}}& \text{ symmetric in } (\scalind{1},\scalind{2})
    \,,\nn
    O^{(6,4)}_{\scalind{1}\scalind{2}\scalind{3}\scalind{4}}& \text{ symmetric in } (\scalind{1},\scalind{2}) \text{ and } (\scalind{3},\scalind{4})
    \,,&
    O^{(6,16)}_{a_1a_2a_3}& \text{ antisymmetric in } (a_2,a_3)
    \,,\nn
    O^{(6,5)}_{\scalind{1}\scalind{2}a_3a_4}& \text{ symmetric in } (\scalind{1},\scalind{2}) \text{ and } (a_3,a_4)
    \,,&
    O^{(6,17)}_{a_1a_2}& \text{ symmetric in } (a_1,a_2)
    \,,\nn
    O^{(6,6)}_{\scalind{1}\scalind{2}a_3a_4}& \text{ no symmetry required}
    \,,&
    O^{(6,18)}_{a_1a_2a_3}& \text{ no symmetry required}
    \,,\nn
    O^{(6,7)}_{\scalind{1}\scalind{2}a_3a_4}& \text{ symmetric in } (\scalind{1},\scalind{2}) \text{ and } (a_3,a_4)
    \,,&
    O^{(6,19)}_{\scalind{1}\scalind{2}a_3a_4}& \text{ no symmetry required}
    \,,\nn
    O^{(6,8)}_{a_1a_2a_3a_4}& \text{ symmetric in } (a_1,a_2) \text{ and } (a_3,a_4)
    \,,\nn
    O^{(6,9)}_{\scalind{1}\scalind{2}\scalind{3}\scalind{4}\scalind{5}\scalind{6}}& \text{ symmetric in } (\scalind{1},\scalind{2},\scalind{3},\scalind{4},\scalind{5},\scalind{6})
    \,,\nn
    O^{(6,10)}_{\scalind{1}\scalind{2}\scalind{3}\scalind{4}a_5a_6}& \text{ symmetric in } (\scalind{1},\scalind{2},\scalind{3},\scalind{4}) \text{ and } (a_5,a_6)
    \,,\nn
    O^{(6,11)}_{\scalind{1}\scalind{2}a_3a_4a_5a_6}& \text{ symmetric in } (\scalind{1},\scalind{2}) \text{ and } (a_3,a_4,a_5,a_6)
    \,,\nn
    O^{(6,12)}_{a_1a_2a_3a_4a_5a_6}& \text{ symmetric in } (a_1,a_2,a_3,a_4,a_5,a_6)
    \,,\nn
    O^{(6,20)}_{a_1a_2a_3a_4}& \text{ symmetric in } (a_1,a_2) \text{ and}
    \nn &\, O^{(6,20)}_{a_1a_2a_3a_4}+O^{(6,20)}_{a_3a_1a_2a_4}+O^{(6,20)}_{a_2a_3a_1a_4}=0
    \,.
\end{align}
Providing tensors that do not respect these symmetries
will cause the determination of the Wilson coefficients to fail.
Note that the mapping from a given tensor to the operator basis
in eq.~\eqref{eq:dim_6_operators} is underdetermined
unless these symmetry constraints are imposed.
The decomposition into the Wilson coefficients is unique only once they are.
As a concrete example, the following schematic script provides
a tensor $O^{(6,12)}_{a_1a_2a_3a_4a_5a_6}$
that is not fully symmetric,
and consequently yields no solution for
$\alpha_{A^6,1}$ and $\alpha_{A^6,2}$:
\begin{lstlisting}[
  language=Mathematica,
  escapeinside={(*@}{@*)}
  ]
Tadj=-I gvvv/gs;
X6=Contract[Tadj,Tadj,Tadj,Tadj,Tadj,Tadj,{{2,18},{3,5},{6,8},{9,11},{12,14},{15,17}}];
O[6,12]=(*@$\alpha$@*)[A^6,1]X6+(*@$\alpha$@*)[A^6,2]TensorProduct[X4,X2];

Dimension6Matching[{O[6,12]},{12},{(*@$\alpha$@*)[A^6,1],(*@$\alpha$@*)[A^6,2]}]
\end{lstlisting}

\subsubsection{Power counting}
\label{sec:power_counting}

The operator matching within
\dralgo{}~\cite{Ekstedt:2022bff} is based on the power-counting scheme
\begin{align}
    \lambda_{\scalind{1}\scalind{2}\scalind{3}\scalind{4}}^{ }
    \sim
    \lambda_{\scalind{1}\scalind{2}\scalind{3}}^{ }
    \sim
    \lambda_{\scalind{1}}^{ } &\sim g^2
    \,,&
    \mu_{\scalind{1}\scalind{2}}^{ } &\sim g^2 T^2
    \,,\nn[1mm]
    Y_{\scalind{1}IJ}^{ } \sim
    g^a_{IJ} \sim
    g^a_{\scalind{1}\scalind{2}} \sim
    g^{abc} &\sim g
    \,,&
    M_{IJ}^{ } &\sim g T
    \,,
\end{align}
where
$g$ is a generic gauge coupling.
The tensors
$\lambda_{\scalind{1}\scalind{2}\scalind{3}\scalind{4}}^{ }$,
$\lambda_{\scalind{1}\scalind{2}\scalind{3}}^{ }$,
$\lambda_{\scalind{1}}^{ }$,
$\mu_{\scalind{1}\scalind{2}}^{ }$,
$M_{IJ}^{ }$,
$Y_{\scalind{1}IJ}^{ }$,
$g^a_{\scalind{1}\scalind{2}}$,
$g^a_{IJ}$,
$g^{abc}$
are defined in the Lagrangian~\eqref{eq:Lag_4d_DRalgo} and
listed in tab.~\ref{tab:coupling_tensors}.
The matching of higher-dimensional operators has been performed
at one-loop order, neglecting zero-temperature mass contributions.
The leading power of the gauge coupling $g$ at which each operator
first appears then varies with the operator type
of the bases in eqs.~\eqref{eq:dim_5_operators} and~\eqref{eq:dim_6_operators},
\begin{align*}
    \mathcal{O}(g^2)&:\quad O^{(6,13)},\,O^{(6,15)},\,O^{(6,17)}\,,\\
    \mathcal{O}(g^3)&:\quad O^{(5,1)},\,O^{(5,5)},\,O^{(5,6)},\,O^{(5,7)},\,O^{(5,8)},\,O^{(6,1)},\,O^{(6,14)},\,O^{(6,16)},\,O^{(6,18)}\,,\\
    \mathcal{O}(g^4)&:\quad O^{(6,2)},\,O^{(6,3)},\,O^{(6,4)},\,O^{(6,5)},\,O^{(6,6)},\,O^{(6,7)},\,O^{(6,8)},\,O^{(6,19)},\,O^{(6,20)}\,,\\
    \mathcal{O}(g^5)&:\quad O^{(5,2)},\,O^{(5,3)},\,O^{(5,4)}\,,\\
    \mathcal{O}(g^6)&:\quad O^{(6,9)},\,O^{(6,10)},\,O^{(6,11)},\,O^{(6,12)}\,.
\end{align*}

\subsubsection{Gauge dependence}
\label{gauge_dependence}

The computation was performed using the background-field gauge~\cite{Abbott:1980hw}, using
the same gauge-fixing parameter {\tt xi} for all the gauge fields.
The corresponding Feynman rules, with the gauge-fixing parameter made explicit and
in Minkowski signature, can be found in~\cite{Schwartz:2014sze}.
See also~\cite{Martin:2018emo} for a generalized gauge fixing.

%
\section{Conclusions}
\label{sec:conclusions}

\begin{table}[t]
  \centering
  \begin{tabular}{|lc|}
  \hline
  Model
  & dim-6 matching  \\
  \hline
  \hline
  Scalar-Yukawa      
  & $10^{-1}$~s \\
  Standard Model
  & \hphantom{$10^{-1}$}\llap{$2045$}~s \\
  Abelian Higgs model
  & \hphantom{$10^{-1}$}\llap{$2.5$}~s \\
  QCD
  & \hphantom{$10^{-1}$}\llap{$729$}~s \\
  \hline
  \end{tabular}
  \caption{Wall-clock timings for the higher-dimensional operator matching
  in \dralgo{} \DRalgoVersion{} for the models discussed in
  sec.~\ref{sec:application_to_models}.
  Timings were obtained on an Apple M3 MacBook Air (8-core, 16\,GB RAM).
  The most time-consuming part of these computations
  is the matching to Wilson coefficients, not
  the tensor contractions for the $O^{(n,k)}$
  which is already taking place during
  {\tt PerformDRhard[]}.
  This is reflected in the
  consistently fast runtimes of {\tt ODIM5} and {\tt ODIM6}.
  Future updates will focus on improving the matching procedure
  at reduced computational cost.
  }
  \label{tab:timings}
\end{table}
Based on the thermal EFT construction software \dralgo{},
we have automated the matching of generic dimension-five and dimension-six
operator bases in thermal high-temperature effective theories.
This development completes the set of matching relations required
to determine all operators up to and including the marginal ones in $3d$~EFTs.

The extension builds upon the existing
coupling tensor framework of~\dralgo{}
and~{\tt GroupMath} for generic gauge theories,
by reusing their representation of coupling tensors.
As a result, the matching of higher-dimensional operators is readily
available for any model that can already be implemented within the package.
We have presented the operator basis employed,
the precise interface of the new functions
{\tt Dimension5Matching} and
{\tt Dimension6Matching}, as well as the auxiliary routines
{\tt ODIM5},
{\tt ODIM6}, and
{\tt HardThermal1LoopInt}.
The required symmetry properties of the input group tensors,
the dependence on the gauge parameter, and
the role of redundant operators have been discussed in detail.

We have validated
the implementation against several known results in the literature:
\begin{itemize}
    \item
      the scalar-Yukawa model~\cite{Chala:2024xll},
    \item
      the electroweak sector of the Standard Model~\cite{Chala:2025aiz}.
      Upon comparison with~\cite{Moore:1995jv}, we find similar discrepancies
      as reported in~\cite{Chala:2025aiz}.
    \item
      the Abelian Higgs model~\cite{Bernardo:2025vkz}, and
    \item
      hot ${\rm SU}(N)$ at fixed $N$
      with fundamental fermions~\cite{Chapman:1994vk,Megias:2003ui,Laine:2018lgj}.
\end{itemize}
For QCD, we have additionally derived the explicit relations
between redundant and physical operator bases,
demonstrating the cancellation of gauge-parameter dependence in the latter.
The fermionic contributions to the dimension-six operator in
EQCD for $\Nc = 3$ have been employed to compute a UV divergence resulting from
the insertion of fermionic dimension-six operators to
the two-loop soft contributions to the magnetostatic wave function
renormalization.
The result is given in~\eqref{eq:deltaZb}.
For the Standard Model,
we have extended previous analyses by computing dimension-six operators
that couple the strong and electroweak sectors, as well as
parity-violating operators, making contact with~\cite{Kajantie:1997ky}.

To give an indication of the computational cost,
tab.~\ref{tab:timings} lists the wall-clock time
required to perform
the dimension-six matching
for each of the models considered in this work,
measured on an Apple M3 MacBook Air (8-core, 16\,GB RAM).

With the full operator basis at this order in hand,
it is now possible to systematically test the validity of
the high-temperature expansion in large-scale phase-transition studies,
where
the assumption of high temperature
is compromised for sufficiently strong transitions~\cite{Chala:2024xll,Bernardo:2026whs}.
Moreover, the automated matching of parity-violating operators opens
the way to
a systematic investigation of CP-odd effects at high temperature
in generic models.
Given that CP violation is a necessary ingredient for baryogenesis,
access to the full set of higher-dimensional CP-violating operators
in thermal effective theories provides a new handle for probing
BSM baryogenesis scenarios in a systematic and model-independent way.

A natural direction for future work is to quantify the convergence
of the higher-dimensional operator expansion for thermodynamic observables
such as the critical temperature or latent heat,
and to benchmark the results against approaches that bypass
the high-temperature expansion
altogether~\cite{Curtin:2022ovx,Chakrabortty:2024wto,Navarrete:2025yxy}.
Deviations between $3d$ and $4d$ lattice simulations have
already been reported for the strongest transitions
in models such as the Standard Model~\cite{Laine:1996nz,Laine:1999rv},
suggesting that the high-temperature approximation may break down
in parts of the parameter space of BSM models.
Further developments include matching
from the soft to the softer scale~\cite{Fuentes-Martin:2026bhr},
extending the computation to two-loop order for
the most phenomenologically relevant higher-dimensional operators, and
implementing automated field redefinitions
(cf.~\cite{Criado:2024mpx,Fonseca:2025zjb}) to obtain results
directly in a non-redundant physical basis.
Finally,
investigating the impact of higher-dimensional operators
in lattice simulations of EQCD and MQCD
could give insight
into the IR behavior of hot QCD~\cite{Laine:2018lgj}.

%
\section*{Acknowledgements}

We thank
Mikael Chala,
Andreas Ekstedt, and
Luis Gil
for enlightening discussions.
We also thank
Tuomas~V.I.\ Tenkanen and
Jorinde van de Vis
for in-depth testing of the code
and thoroughly commenting on the manuscript.
FB and PS were supported by
the Swiss National Science Foundation (SNSF) under grant
\href{https://data.snf.ch/grants/grant/215997}{\tt PZ00P2-215997}. 

%
\subsubsection*{Data availability statement}

The version of \dralgo{} corresponding
to this publication is \DRalgoVersion{},
which is available at
\url{https://github.com/DR-algo/DRalgo}.
With this version,
we also provide example model files demonstrating
matching to higher-dimensional operators:
\begin{itemize}
    \item
    Scalar-Yukawa in sec.~\ref{sec:firstExample}:\\
    \verb|https://github.com/DR-algo/DRalgo/examples/ScalYukawa_HDO.m|,
    \item
    QCD in sec.~\ref{sec:QCD}:\\
    \verb|https://github.com/DR-algo/DRalgo/examples/QCD_HDO.m|,
    \item
    Standard Model in sec.~\ref{sec:SM}:\\
    \verb|https://github.com/DR-algo/DRalgo/examples/sm_HDO.m|,
    \item
    Abelian-Higgs model in~\cite{Bernardo:2025vkz}:\\
    \verb|https://github.com/DR-algo/DRalgo/examples/ah_HDO.m|,
    \item
    $SU(2)+\text{Higgs}$ in~\cite{Chala:2025aiz}:\\
    \verb|https://github.com/DR-algo/DRalgo/examples/SU2+Higgs_HDO.m|.
\end{itemize}

\appendix
\renewcommand{\thesection}{\Alph{section}}
\renewcommand{\thesubsection}{\Alph{section}.\arabic{subsection}}
\renewcommand{\theequation}{\Alph{section}.\arabic{equation}}

\section{Standard Model higher-dimensional operators}
\label{appendix:SM}

In the Standard Model,
the complete set of operators
arising in dimensional reduction at dimension-six level is
\begin{align}
    \label{eq:dim6_SM}
    \mathcal{L}^{(6,1)}_\rmii{SM}&=
        \alpha_{W^3}\,\epsilon^{\suii{1}\suii{2}\suii{3}}W^{\suii{1}}_{ij}W^{\suii{2}}_{jk}W^{\suii{3}}_{jk}
      + \alpha_{G^3}\,f^{\suiii{1}\suiii{2}\suiii{3}}G^{\suiii{1}}_{ij}G^{\suiii{2}}_{jk}G^{\suiii{3}}_{jk}
    \,,\nn
    \mathcal{L}^{(6,2)}_\rmii{SM}&=
        \alpha_{H^2 B^2}B_{ij}B_{ij}|H|^2
      + \alpha_{H^2 W^2}W^{\suii{1}}_{ij}W^{\suii{1}}_{ij}|H|^2
    \nn &
      + \alpha_{H^2 B W}B_{ij}W^{\suii{1}}_{ij}H^\dagger\sigma^{\suii{1}}H
      + \alpha_{H^2G^2}G^{\suiii{1}}_{ij}G^{\suiii{1}}_{ij}|H|^2
    \,,\nn
    \mathcal{L}^{(6,3)}_\rmii{SM}&=
        \alpha_{B_0^2 B^2}B_{ij}^2B_0^2
      + \alpha_{W_0^2 W^2,1}W^{\suii{1}}_{ij}W^{\suii{1}}_{ij}W^{\suii{2}}_0W^{\suii{2}}_0
      + \alpha_{W_0^2 W^2,2}W^{\suii{1}}_{ij}W^{\suii{2}}_{ij}W^{\suii{1}}_0W^{\suii{2}}_0
    \nn &
      + \alpha_{B_0^2 W^2}W^{\suii{1}}_{ij}W^{\suii{1}}_{ij}B_0^2
      + \alpha_{B_0 W_0 B\, W}W^{\suii{1}}_{ij}B_{ij}B_0 W^{\suii{1}}_0
      + \alpha_{W_0^2B^2}B_{ij}^2W_0^{\suii{1}}W_0^{\suii{1}}
    \nn &
      + \alpha_{G_0^2G^2,1}G^{\suiii{1}}_{ij}G^{\suiii{1}}_{ij}G^{\suiii{2}}_0G^{\suiii{2}}_0
      + \alpha_{G_0^2G^2,2}G^{\suiii{1}}_{ij}G^{\suiii{2}}_{ij}G^{\suiii{1}}_0G^{\suiii{2}}_0
    \nn &
      + \alpha_{G_0^2G^2,3}X^{\suiii{1}\suiii{3}\suiii{2}\suiii{4}}G_{ij}^{\suiii{1}}G_{ij}^{\suiii{2}}G_0^{\suiii{3}}G_0^{\suiii{4}}
      + \alpha_{G_0^2G B}d^{\suiii{1}\suiii{2}\suiii{3}}G^{\suiii{1}}_{ij}B_{ij}G^{\suiii{2}}_0G^{\suiii{3}}_0
    \nn &
      + \alpha_{G_0 B_0 G^2}\,d^{\suiii{1}\suiii{2}\suiii{3}}G^{\suiii{1}}_{ij}G^{\suiii{2}}_{ij}G^{\suiii{3}}_0B_0
      + \alpha_{G_0^2B^2}B_{ij}^2G^{\suiii{1}}_0G^{\suiii{1}}_0
      + \alpha_{G_0B_0 G\,B}B_{ij}G^{\suiii{1}}_{ij}G^{\suiii{1}}_0B_0
    \nn &
      + \alpha_{B_0^2 G^2}G^{\suiii{1}}_{ij}G^{\suiii{1}}_{ij}B_0^2
      + \alpha_{G_0^2W^2}W^{\suii{1}}_{ij}W^{\suii{1}}_{ij}G^{\suiii{1}}_0G^{\suiii{1}}_0
      + \alpha_{G_0W_0\,G^2}W^{\suii{1}}_{ij}G^{\suiii{1}}_{ij}W^{\suii{1}}_0G^{\suiii{1}}_0
    \nn &
      + \alpha_{W_0^2G^2}G^{\suiii{1}}_{ij}G^{\suiii{1}}_{ij}W^{\suii{1}}_0W^{\suii{1}}_0
    \,,\nn
    \mathcal{L}^{(6,4)}_\rmii{SM}&=
        \alpha_{H^4D^2,1}\re\{J_i^2\}
      + \alpha_{H^4D^2,2}\im\{J_i^2\}
      + \alpha_{H^4D^2,3}|J_i|^2
      + \alpha_{H^4D^2,4}|D_iH|^2|H|^2
    \,,\nn
    \mathcal{L}^{(6,5)}_\rmii{SM}&=
        \alpha_{H^2 B_0^2 D^2,1}(D_iH^\dagger D_iH)B_0^2
      + \alpha_{H^2 W_0^2 D^2,1}(D_iH^\dagger D_iH)W^{\suii{1}}_0W^{\suii{1}}_0
    \nn &
      + \alpha_{H^2 B_0 W_0 D^2,1}(D_iH^\dagger \sigma^{\suii{1}} D_iH)W^{\suii{1}}_0B_0
      + \alpha_{H^2 G_0^2 D^2,1}(D_iH^\dagger D_iH)G^{\suiii{1}}_0G^{\suiii{1}}_0
    \,,\nn
    \mathcal{L}^{(6,6)}_\rmii{SM}&=
        \alpha_{H^2B_0^2D^2,2}\re\{J_i\}(D_iB_0)B_0
      + \alpha_{H^2B_0^2D^2,3}\im\{J_i\}(D_iB_0)B_0
    \nn &
      + \alpha_{H^2W_0^2D^2,2}\re\{J_i\}(D_iW_0^{\suii{1}})W_0^{\suii{1}}
      + \alpha_{H^2W_0^2D^2,3}\im\{J_i\}(D_iW_0^{\suii{1}})W_0^{\suii{1}}
    \nn &
      + \alpha_{H^2W_0^2D^2,4}\epsilon^{\suii{1}\suii{2}\suii{3}}\re\{J^{\suii{1}}_i\}(D_iW_0^{\suii{2}})W_0^{\suii{3}}
      + \alpha_{H^2W_0^2D^2,5}\epsilon^{\suii{1}\suii{2}\suii{3}}\im\{J^{\suii{1}}_i\}(D_iW_0^{\suii{2}})W_0^{\suii{3}}
    \nn &
      + \alpha_{H^2B_0W_0D^2,2}\re\{J^{\suii{1}}_i\}(D_iW_0^{\suii{1}})B_0
      + \alpha_{H^2B_0W_0D^2,3}\im\{J^{\suii{1}}_i\}(D_iW_0^{\suii{1}})B_0
    \nn &
      + \alpha_{H^2B_0W_0D^2,4}\re\{J^{\suii{1}}_i\}(D_iB_0)W_0^{\suii{1}}
      + \alpha_{H^2B_0W_0D^2,5}\im\{J^{\suii{1}}_i\}(D_iB_0)W_0^{\suii{1}}
    \nn &
      + \alpha_{H^2G_0^2D^2,2}\re\{J_i\}(D_iG_0^{\suiii{1}})G_0^{\suiii{1}}
      + \alpha_{H^2G_0^2D^2,3}\im\{J_i\}(D_iG_0^{\suiii{1}})G_0^{\suiii{1}}
    \,,\nn
    \mathcal{L}^{(6,7)}_\rmii{SM}&=
        \alpha_{H^2 B_0^2 D^2,4}|H|^2(D_iB_0)^2
      + \alpha_{H^2 W_0^2 D^2,6}|H|^2(D_iW_0^{\suii{1}})(D_iW_0^{\suii{1}})
    \nn &
      + \alpha_{H^2 W_0 B_0 D^2,6}(H^\dagger\sigma^{\suii{1}}H)(D_iW_0^{\suii{1}})(D_iB_0)
      + \alpha_{H^2 G_0^2 D^2,4}|H|^2(D_iG_0^{\suiii{1}})(D_iG_0^{\suiii{1}})
    \,,\nn
    \mathcal{L}^{(6,8)}_\rmii{SM}&=
        \alpha_{B_0^4D^2}(D_iB_0)^2B_0^2
      + \alpha_{B_0^2W_0^2D^2,1}(D_iB_0)^2W_0^{\suii{1}}W_0^{\suii{1}}
    \nn &
      + \alpha_{B_0^2W_0^2D^2,2}(D_iB_0)(D_iW_0^{\suii{1}})W_0^{\suii{1}}B_0
      + \alpha_{B_0^2W_0^2D^2,3}D_iW^{\suii{1}}_0D_iW^{\suii{1}}_0B_0^2
    \nn &
      + \alpha_{W_0^4D^2,1}(D_iW_0^{\suii{1}}D_iW_0^{\suii{1}})W_0^{\suii{2}}W_0^{\suii{2}}
      + \alpha_{W_0^4D^2,2}(D_iW_0^{\suii{1}}D_iW_0^{\suii{2}})W_0^{\suii{1}}W_0^{\suii{2}}
    \nn &
      + \alpha_{G_0^4D^2,1}(D_iG_0^{\suiii{1}}D_iG_0^{\suiii{1}})G_0^{\suiii{2}}G_0^{\suiii{2}}
      + \alpha_{G_0^4D^2,2}(D_iG_0^{\suiii{1}}D_iG_0^{\suiii{2}})G_0^{\suiii{1}}G_0^{\suiii{2}}
    \nn &
      + \alpha_{G_0^4D^2,3}X^{\suiii{1}\suiii{3}\suiii{2}\suiii{4}}(D_iG_0^{\suiii{1}}D_iG_0^{\suiii{2}})G_0^{\suiii{3}}G_0^{\suiii{4}}
      + \alpha_{G_0^3B_0D^2,1}d^{\suiii{1}\suiii{2}\suiii{3}}(D_iB_0\,D_iG_0^{\suiii{1}})G_0^{\suiii{2}}G_0^{\suiii{3}}
    \nn &
      + \alpha_{G_0^3B_0D^2,2}d^{\suiii{1}\suiii{2}\suiii{3}}(D_iG_0^{\suiii{1}}D_iG_0^{\suiii{2}})G_0^{\suiii{3}}B_0
      + \alpha_{G_0^2B_0^2D^2,1}(D_iB_0)^2G_0^{\suiii{1}}G_0^{\suiii{1}}
    \nn &
      + \alpha_{G_0^2B_0^2D^2,2}(D_iB_0\,D_iG_0^{\suiii{1}})G_0^{\suiii{1}}B_0
      + \alpha_{G_0^2B_0^2D^2,3}(D_iG_0^{\suiii{1}}D_iG_0^{\suiii{1}})B_0^2
    \nn &
      + \alpha_{G_0^2W_0^2D^2,1}(D_iW_0^{\suii{1}}D_iW_0^{\suii{1}})G_0^{\suiii{1}}G_0^{\suiii{1}}
      + \alpha_{G_0^2W_0^2D^2,2}(D_iG_0^{\suiii{1}}D_iW_0^{\suii{1}})W_0^{\suii{1}}G_0^{\suiii{1}}
    \nn &
      + \alpha_{G_0^2W_0^2D^2,3}(D_iG_0^{\suiii{1}}D_iG_0^{\suiii{1}})W_0^{\suii{1}}W_0^{\suii{1}}
    \,,\nn
    \mathcal{L}^{(6,9)}_\rmii{SM}&=
        \alpha_{H^6}|H|^6
    \,,\nn[1mm]
    \mathcal{L}^{(6,10)}_\rmii{SM}&=
        \alpha_{H^4 B_0^2}|H|^4B_0^2
      + \alpha_{H^4 W_0^2,1}|H|^4W^{\suii{1}}_0W^{\suii{1}}_0
      + \alpha_{H^4 W_0^2,2}(H^\dagger\sigma^{\suii{1}}H)(H^\dagger\sigma^{\suii{2}}H)W^{\suii{1}}_0W^{\suii{2}}_0
    \nn &
      + \alpha_{H^4 B_0 W_0}|H|^2(H^\dagger\sigma^{\suii{1}}H)W^{\suii{1}}_0B_0
      + \alpha_{H^4 G_0^2}|H|^4G^{\suiii{1}}_0G^{\suiii{1}}_0
    \,,\nn
    \mathcal{L}^{(6,11)}_\rmii{SM}&=
        \alpha_{H^2B_0^4}|H|^2B_0^4
      + \alpha_{H^2B_0^3W_0}(H^\dagger\sigma^{\suii{1}}H)W^{\suii{1}}_0B_0^3
      + \alpha_{H^2 B_0^2 W_0^2}|H|^2W^{\suii{1}}_0W^{\suii{1}}_0B_0^2
    \nn &
      + \alpha_{H^2 B_0 W_0^3}(H^\dagger\sigma^{\suii{1}}H)W^{\suii{1}}_0W^{\suii{2}}_0W^{\suii{2}}_0B_0
      + \alpha_{H^2 W_0^4}|H|^2W^{\suii{1}}_0W^{\suii{1}}_0W^{\suii{2}}_0W^{\suii{2}}_0
    \nn &
      + \alpha_{H^2 G_0^2 B_0^2}|H|^2G^{\suiii{1}}_0G^{\suiii{1}}_0B_0^2
      + \alpha_{H^2 G_0^2 W_0^2}|H|^2G^{\suiii{1}}_0G^{\suiii{1}}_0W^{\suii{1}}_0W^{\suii{1}}_0
    \nn &
      + \alpha_{H^2 G_0^2 W_0 B_0}(H^\dagger\sigma^{\suii{1}}H)W_0^{\suii{1}}G^{\suiii{1}}_0G^{\suiii{1}}_0B_0
      + \alpha_{H^2 G_0^3 B_0}d^{\suiii{1}\suiii{2}\suiii{3}}|H|^2G_0^{\suiii{1}}G_0^{\suiii{2}}G_0^{\suiii{3}}B_0
    \nn &
      + \alpha_{H^2 G_0^3 W_0}d^{\suiii{1}\suiii{2}\suiii{3}}(H^\dagger\sigma^{\suii{1}}H)W^{\suii{1}}_0G_0^{\suiii{1}}G_0^{\suiii{2}}G_0^{\suiii{3}}
      + \alpha_{H^2G_0^4}|H|^2G_0^{\suiii{1}}G_0^{\suiii{1}}G_0^{\suiii{2}}G_0^{\suiii{2}}
    \,,\nn[1mm]
    \mathcal{L}^{(6,12)}_\rmii{SM}&=
        \alpha_{B_0^6}B_0^6
      + \alpha_{B_0^4W_0^2}B_0^4W_0^{\suii{1}}W_0^{\suii{1}}
      + \alpha_{B_0^2W_0^4}B_0^2W_0^{\suii{1}}W_0^{\suii{1}}W_0^{\suii{2}}W_0^{\suii{2}}
    \nn &
      + \alpha_{W_0^6}W_0^{\suii{1}}W_0^{\suii{1}}W_0^{\suii{2}}W_0^{\suii{2}}W_0^{\suii{3}}W_0^{\suii{3}}
      + \alpha_{G_0^2B_0^4}G_0^{\suiii{1}}G_0^{\suiii{1}}B_0^4
      + \alpha_{G_0^2B_0^2W_0^2}G_0^{\suiii{1}}G_0^{\suiii{1}}W_0^{\suii{1}}W_0^{\suii{1}}B_0^2
    \nn &
      + \alpha_{G_0^2W_0^4}G_0^{\suiii{1}}G_0^{\suiii{1}}W_0^{\suii{1}}W_0^{\suii{1}}W_0^{\suii{2}}W_0^{\suii{2}}
      + \alpha_{G_0^3B_0^3}d^{\suiii{1}\suiii{2}\suiii{3}}G_0^{\suiii{1}}G_0^{\suiii{2}}G_0^{\suiii{3}}B_0^3
    \nn &
      + \alpha_{G_0^3B_0 W_0^2}d^{\suiii{1}\suiii{2}\suiii{3}}G_0^{\suiii{1}}G_0^{\suiii{2}}G_0^{\suiii{3}}B_0 W_0^{\suii{1}}W_0^{\suii{1}}
      + \alpha_{G_0^4B_0^2}G_0^{\suiii{1}}G_0^{\suiii{1}}G_0^{\suiii{2}}G_0^{\suiii{2}}B_0^2
    \nn &
      + \alpha_{G_0^4W_0^2}G_0^{\suiii{1}}G_0^{\suiii{1}}G_0^{\suiii{2}}G_0^{\suiii{2}}W_0^{\suii{1}}W_0^{\suii{1}}
      + \alpha_{G_0^5B_0}d^{\suiii{1}\suiii{2}\suiii{3}}G_0^{\suiii{1}}G_0^{\suiii{2}}G_0^{\suiii{3}}G_0^{\suiii{4}}G_0^{\suiii{4}}B_0
    \nn &
      + \alpha_{G_0^6,1}d^{\suiii{1}\suiii{2}\suiii{3}}d^{\suiii{4}\suiii{5}\suiii{6}}G_0^{\suiii{1}}G_0^{\suiii{2}}G_0^{\suiii{3}}G_0^{\suiii{4}}G_0^{\suiii{5}}G_0^{\suiii{6}}
      + \alpha_{G_0^6,2}G_0^{\suiii{1}}G_0^{\suiii{1}}G_0^{\suiii{2}}G_0^{\suiii{2}}G_0^{\suiii{3}}G_0^{\suiii{3}}
    \,,\nn[1mm]
    \mathcal{L}^{(6,13)}_\rmii{SM}&=
        \alpha_{W^2D^2}(D_iW_{ij}^{\suii{1}})(D_kW_{kj}^{\suii{1}})
      + \alpha_{B^2D^2}(D_iB_{ij})(D_kB_{kj})
      + \alpha_{G^2D^2}(D_iG_{ij}^{\suiii{1}})(D_kG_{kj}^{\suiii{1}})
    \,,\nn[1mm]
    \mathcal{L}^{(6,14)}_\rmii{SM}&=
        \alpha_{H^2 W D^2}(D_j W_{ij}^{\suii{1}})\im\{J^{\suii{1}}_i\}
      + \alpha_{H^2 B D^2}(D_j B_{ij})\im\{J_i\}
    \,,\nn[1mm]
    \mathcal{L}^{(6,15)}_\rmii{SM}&=
        \alpha_{H^2D^4}D^2H^\dagger D^2H
    \,,\nn[1mm]
    \mathcal{L}^{(6,16)}_\rmii{SM}&=
        \alpha_{W_0^2 W D^2}\epsilon^{\suii{1}\suii{2}\suii{3}}(D_j W_{ij}^{\suii{1}})(D_iW_0^{\suii{2}}W_0^{\suii{3}})
      + \alpha_{W_0 B_0 W D^2}(D_j W_{ij}^{\suii{1}})(D_iW_0^{\suii{1}}B_0-W_0^{\suii{1}}D_iB_0)
    \nn &
      + \alpha_{G_0^2 G D^2}f^{\suiii{1}\suiii{2}\suiii{3}}(D_jG_{ij}^{\suiii{1}})(D_iG_0^{\suiii{2}}G_0^{\suiii{3}})
      + \alpha_{G_0 B_0 G D^2}(D_j G_{ij}^{\suiii{1}})(D_iG_0^{\suiii{1}}B_0-G_0^{\suiii{1}}D_iB_0)
    \,,\nn
    \mathcal{L}^{(6,17)}_\rmii{SM}&=
        \alpha_{W_0^2D^4}D^2W_0^{\suii{1}} D^2W_0^{\suii{1}}
      + \alpha_{B_0^2D^4}D^2B_0 D^2B_0
      + \alpha_{G_0^2D^4}D^2G_0^{\suiii{1}} D^2G_0^{\suiii{1}}
    \,,\nn
    \mathcal{L}^{(6,18)}_\rmii{SM}&=
        i\alpha_{B_0 B^2 D}B_0 \widetilde{B}_k (D_l B_{lk})
      + i\alpha_{B_0 W^2 D}B_0 \widetilde{W}^{\suii{1}}_k (D_l W^{\suii{1}}_{lk})
      + i\alpha_{W_0 B\,W\,D,1}W^{\suii{1}}_0 \widetilde{B}_k (D_l W^{\suii{1}}_{lk})
    \nn &
      + i\alpha_{W_0 B\,W\,D,2}W^{\suii{1}}_0 \widetilde{W}^{\suii{1}}_k (D_l B_{lk})
      + i\alpha_{W_0 W^2\,D}\epsilon^{\suii{1}\suii{2}\suii{3}}W^{\suii{1}}_0\widetilde{W}^{\suii{2}}_k (D_l W^{\suii{3}}_{lk})
    \nn &
      + i\alpha_{B_0 G^2 D}B_0 \widetilde{G}^{\suiii{1}}_k (D_l G^{\suiii{1}}_{lk})
      + i\alpha_{G_0 B\,G\,D,1}G^{\suiii{1}}_0 \widetilde{B}_k (D_l G^{\suiii{1}}_{lk})
      + i\alpha_{G_0 B\,G\,D,2}G^{\suiii{1}}_0 \widetilde{G}^{\suiii{1}}_k (D_l B_{lk})
    \nn &
      + i\alpha_{G_0 G^2\,D,1}f^{\suiii{1}\suiii{2}\suiii{3}}G^{\suiii{1}}_0\widetilde{G}^{\suiii{2}}_k (D_l G^{\suiii{3}}_{lk})
      + i\alpha_{G_0 G^2\,D,2}d^{\suiii{1}\suiii{2}\suiii{3}}G^{\suiii{1}}_0\widetilde{G}^{\suiii{2}}_k (D_l G^{\suiii{3}}_{lk})
    \,,\nn
    \mathcal{L}^{(6,19)}_\rmii{SM}&=
        i\alpha_{H^2 B_0 B\,D,1}\re\{J_i\}B_0\widetilde{B}_i
      + i\alpha_{H^2 B_0 B\,D,2}\im\{J_i\}B_0\widetilde{B}_i
      + i\alpha_{H^2 B_0 W\,D,1}\re\{J^{\suii{1}}_i\}B_0\widetilde{W}^{\suii{1}}_i
    \nn &
      + i\alpha_{H^2 B_0 W\,D,2}\im\{J^{\suii{1}}_i\}B_0\widetilde{W}^{\suii{1}}_i
      + i\alpha_{H^2 W_0 B\,D,1}\re\{J^{\suii{1}}_i\}W_0^{\suii{1}}\widetilde{B}_i
    \nn &
      + i\alpha_{H^2 W_0 B\,D,2}\im\{J^{\suii{1}}_i\}W_0^{\suii{1}}\widetilde{B}_i
      + i\alpha_{H^2 W_0 W\,D,1}\re\{J_i\}W_0^{\suii{1}}\widetilde{W}_i^{\suii{1}}
    \nn &
      + i\alpha_{H^2 W_0 W\,D,2}\im\{J_i\}W_0^{\suii{1}}\widetilde{W}_i^{\suii{1}}
      + i\alpha_{H^2 W_0 W\,D,3}\epsilon^{\suii{1}\suii{2}\suii{3}}\re\{J^{\suii{1}}_i\}W_0^{\suii{2}}\widetilde{W}_i^{\suii{3}}
    \nn &
      + i\alpha_{H^2 W_0 W\,D,4}\epsilon^{\suii{1}\suii{2}\suii{3}}\im\{J^{\suii{1}}_i\}W_0^{\suii{2}}\widetilde{W}_i^{\suii{3}}
    \nn &
      + i\alpha_{H^2 G_0 G\,D,1}\re\{J_i\}G_0^{\suiii{1}}\widetilde{G}_i^{\suiii{1}}
      + i\alpha_{H^2 G_0 G\,D,2} \im\{J_i\}G_0^{\suiii{1}}\widetilde{G}_i^{\suiii{1}}
    \,,\nn
    \mathcal{L}^{(6,20)}_\rmii{SM}&=
        i\alpha_{B_0^2W_0 W\,D}B_0^2(D_iW_0)^{\suii{1}}\widetilde{W}_i^{\suii{1}}
      + i\alpha_{B_0W_0^2 B\,D}B_0W_0^{\suii{1}}(D_iW_0)^{\suii{1}}\widetilde{B}_i
    \nn &
      + i\alpha_{B_0W_0^2W\,D}\epsilon^{\suii{1}\suii{2}\suii{3}}B_0W_0^{\suii{1}}(D_iW_0)^{\suii{2}}\widetilde{W}_i^{\suii{3}}
      + i\alpha_{W_0^3W\,D}W_0^{\suii{1}}W_0^{\suii{1}}(D_iW_0)^{\suii{2}}\widetilde{W}^{\suii{2}}_i
    \nn &
      + i\alpha_{B_0^2G_0 G\,D}B_0^2(D_iG_0)^{\suiii{1}}\widetilde{G}_i^{\suiii{1}}
      + i\alpha_{B_0G_0^2 G\,D,1}B_0G_0^{\suiii{1}}(D_iG_0)^{\suiii{1}}\widetilde{B}_i
    \nn &
      + i\alpha_{B_0G_0^2G\,D,2}f^{\suiii{1}\suiii{2}\suiii{3}}B_0G_0^{\suiii{1}}(D_iG_0)^{\suiii{2}}\widetilde{G}_i^{\suiii{3}}
      + i\alpha_{G_0^3G\,D}G_0^{\suiii{1}}G_0^{\suiii{1}}(D_iG_0)^{\suiii{2}}\widetilde{G}^{\suiii{2}}_i
    \nn &
      + i\alpha_{B_0G_0^2G\,D,3}d^{\suiii{1}\suiii{2}\suiii{3}}G_0^{\suiii{1}}G_0^{\suiii{2}}(D_iB_0)\widetilde{G}_i^{\suiii{3}}
      + i\alpha_{W_0^2G_0 G\, D}W_0^{\suii{1}}W_0^{\suii{1}}(D_iG_0^{\suiii{1}})\widetilde{G}_i^{\suiii{1}}
    \nn &
      + i\alpha_{G_0^2 W_0 W\,D}G_0^{\suiii{1}}G_0^{\suiii{1}}(D_iW_0^{\suii{1}})\widetilde{W}^{\suii{1}}_i
    \,,
\end{align}
where
$B$ denotes the ${\rm U}(1)_\rmii{$Y$}$ gauge field,
$W^{\suii{i}}$ the ${\rm SU}(2)_\rmii{$L$}$ gauge fields with adjoint indices $\suii{i}$,
and $G^{\suiii{i}}$ the ${\rm SU}(3)_\rmi{c}$ gluon fields with color indices $\suiii{i}$;
$B_{ij}$, $W_{ij}$, $G_{ij}$ denote the corresponding field-strength tensors.
The fields
with a subscript $0$ denote the temporal components of the gauge fields.
The tensors $X^{\suiii{1} \suiii{2} \dots \suiii{n}}$ denote
the trace of
a product of ${\rm SU}(3)$ adjoint-representation generators,
\begin{align}
    X^{\suiii{1}\suiii{2}\dots\suiii{n}}&\equiv
      \tr\Bigl\{
        T_{{\rm SU}(3)\,\text{adj}}^{\suiii{1}}
        T_{{\rm SU}(3)\,\text{adj}}^{\suiii{2}}
        \dots
        T_{{\rm SU}(3)\,\text{adj}}^{\suiii{n}}
      \Bigr\}
    \,,
\end{align}
and $J_i$, $J^{\suii{1}}_i$ are covariant-derivative Higgs bilinears,
with and without an ${\rm SU}(2)_\rmii{$L$}$ generator insertion, respectively,
\begin{align}
    J_i&\equiv H^\dagger D_iH
    \,,&
    J^{\suii{1}}_i&\equiv H^\dagger \sigma^{\suii{1}}D_iH
    \,.
\end{align}
We have generated the operator basis with the help of
{\tt BasisGen}~\cite{Criado:2019ugp}.
The expression of the Wilson coefficients can be found in the ancillary file
\verb|SM_HDO.txt|.

\section{Hot QCD operator basis and field redefinitions}
\label{appendix:QCD}

The operator basis used in~\cite{Laine:2018lgj}
to study higher-dimensional operators in finite-temperature QCD
differs from the one introduced above.
It reads
\begin{align}
    \mathcal{L}_{6,\rmii{QCD}}^{(3d)}&=
			\gE^2\,\tr\Big\{c_1\left(D_\mu F_{\mu\nu}\right)^2+c_2\left(D_\mu F_{\mu0}\right)^2
		\nn &\phantom{{}=\,\gE^2\tr\Big\{}
		+ i\gE\bigl[
				c_3\,F_{\mu\nu}F_{\nu\rho}F_{\rho\mu}
			+ c_4\,F_{0\mu}F_{\mu\nu}F_{\nu0}
			+ c_5\,A_0(D_\mu F_{\mu\nu})F_{0\nu}
		\bigr]
		\nn &\phantom{{}=\,\gE^2\tr\Big\{}
		+ \gE^2\bigl[
				c_6\, A_0^2F_{\mu\nu}^2
			+ c_7\, A_0F_{\mu\nu}A_0 F_{\mu\nu}
			+ c_8\,A_0^2F_{0\mu}^2+c_9\,A_0F_{0\mu}A_0F_{0\mu}
			\bigr]
		\nn &\phantom{{}=\,\gE^2\tr\Big\{}
		+ \gE^4\bigl[c_{10}\,A_0^6\bigr]
		\Big\}
		\,,
\end{align}
and can be related to that in eq.~\eqref{eq:off_shell_QCD_lag} through
\begin{align}
    \alpha_{D^2F^2}&=2\,\gE^2\,c_1
		\,,&
    \alpha_{D^4A_0^2}&=2\,\gE^2\,(c_1+c_2)
		\,,\nn[1mm]
    \alpha_{F^3}&=3!\,i\,\gE^3\,c_3
		\,,&
    \alpha_{D^2A_0^2F}&=2i\,\gE^3\,\Bigl(
          2c_1-\frac{3}{2}c_3
        - \frac{1}{2}c_4
        + \frac{1}{2}c_5
        \Bigr)
		\,,\nn[1mm]
    \alpha_{A_0^2F^2,1}&=2!2!\,\gE^4\Bigl(\frac{3}{2}c_3+\frac{1}{2}c_4+c_6\Bigr)
		\,,&
    \alpha_{A_0^2F^2,2}&=2!2!\,\gE^4\Bigl(-\frac{3}{2}c_3-\frac{1}{2}c_4+c_7\Bigr)
		\,,\nn[1mm]
    \alpha_{D^2A_0^4,1}&=2!2!\,\gE^4\Bigl(2c_1+c_5+2c_6+c_8\Bigr)
		\,,&
    \alpha_{D^2A_0^4,2}&=2!2!\,\gE^4\Bigl(-2c_1-c_5+2c_7+c_9\Bigr)
		\,,\nn[1mm]
    \alpha_{A_0^6,1}&=6!\,\gE^6\,c_{10}
		\,.
\end{align}

The operator basis defined in eq.~\eqref{eq:off_shell_QCD_lag} is redundant,
since appropriate field redefinitions of the gauge fields $A_i$ and $A_0$
allow certain interactions to be eliminated in favor of shifts in
the Wilson coefficients of the physical operators.
The subset of operators that cannot be removed in this way defines the physical basis,
given in eq.~\eqref{eq:on_shell_QCD_lag}.
Using {\tt Matchete}~\cite{Fuentes-Martin:2022jrf},
we derive the relations between the coefficients $\beta_i$ and $\alpha_i$:
\begin{align}
\label{eq:QCD_FR}
    \beta_{F^3}&=\alpha_{F^3}
    \,,& 
    \beta_{A_0^2F^2,1}&=
        \alpha_{A_0^2F^2,1}
    \,,& 
    \beta_{A_0^2F^2,2}&=
        \alpha_{A_0^2F^2,2}
    \,,\nn
    \beta_{D^2A_0^4,2}&=
        \alpha_{D^2A_0^4,2}
    \,,& 
    \beta_{A_0^6,1}&=
        \alpha_{A_0^6,1}
    \,,& 
    \beta_{A_0^6,2}&=
        \alpha_{A_0^6,2}
    \,,
\end{align}
with only one non-trivial relation for the coefficient of
the operator $D^2A_0^4$
in eq.~\eqref{eq:on_shell_QCD_lag},
\begin{align}
    \beta_{D^2A_0^4,1}&=
          \alpha_{D^2A_0^4,2}
        - \frac{1}{2}\alpha_{D^2A_0^4,1}
        - 6i\,\gs\,\alpha_{D^2A_0^2F}
        - 6\,\gs^2\,\alpha_{D^2F^2}
    \,.
\end{align}

%
{\small 
    \bibliographystyle{utphys}
    \bibliography{ref}
}
\end{document}